\documentstyle[epsfig,psfig]{mn}
\newif\ifAMStwofonts
\AMStwofontstrue

\def\.{\'\i}
\def\noi{\noindent}
\def\etal{{\it et al. }}
\newcommand{\Hi}{\hbox{H\,{\sc i}}\ }
\newcommand{\Hii}{\hbox{H\,{\sc ii}}\ }

\newcommand{\Nii}{[\hbox{N\,{\sc ii}}]\ }
\newcommand{\Oiii}{[\hbox{O\,{\sc iii}}]\ }

\ifoldfss
  \ifCUPmtlplainloaded \else
    \NewTextAlphabet{textbfit} {cmbxti10} {}
    \NewTextAlphabet{textbfss} {cmssbx10} {}
    \NewMathAlphabet{mathbfit} {cmbxti10} {} 
    \NewMathAlphabet{mathbfss} {cmssbx10} {} 
  \fi
  \ifAMStwofonts
    \ifCUPmtlplainloaded \else
      \NewSymbolFont{upmath} {eurm10}
      \NewSymbolFont{AMSa} {msam10}
      \NewMathSymbol{\upi}     {0}{upmath}{19}
      \NewMathSymbol{\umu}     {0}{upmath}{16}
      \NewMathSymbol{\upartial}{0}{upmath}{40}
      \NewMathSymbol{\leqslant}{3}{AMSa}{36}
      \NewMathSymbol{\geqslant}{3}{AMSa}{3E}

      \let\leq=\leqslant 
      \let\geq=\geqslant 
    \fi
  \fi
\fi 

\ifnfssone
  \newmathalphabet{\mathit}
  \addtoversion{normal}{\mathit}{cmr}{m}{it}
  \addtoversion{bold}{\mathit}{cmr}{bx}{it}
  \newmathalphabet{\mathbfit} 
  \addtoversion{normal}{\mathbfit}{cmr}{bx}{it}
  \addtoversion{bold}{\mathbfit}{cmr}{bx}{it}
  \newmathalphabet{\mathbfss} 
  \addtoversion{normal}{\mathbfss}{cmss}{bx}{n}
  \addtoversion{bold}{\mathbfss}{cmss}{bx}{n}
  \ifAMStwofonts
    \ifCUPmtlplainloaded \else
      \UseAMStwoboldmath
      \makeatletter
      \new@mathgroup\upmath@group
      \define@mathgroup\mv@normal\upmath@group{eur}{m}{n}
      \define@mathgroup\mv@bold\upmath@group{eur}{b}{n}
      \edef\UPM{\hexnumber\upmath@group}
      \new@mathgroup\amsa@group
      \define@mathgroup\mv@normal\amsa@group{msa}{m}{n}
      \define@mathgroup\mv@bold\amsa@group{msa}{m}{n}
      \edef\AMSa{\hexnumber\amsa@group}
      \makeatother
      \mathchardef\upi="0\UPM19
      \mathchardef\umu="0\UPM16
      \mathchardef\upartial="0\UPM40
      \mathchardef\leqslant="3\AMSa36
      \mathchardef\geqslant="3\AMSa3E

      \let\leq=\leqslant 
      \let\geq=\geqslant 
    \fi
  \fi
\fi 

\ifnfsstwo
  \DeclareMathAlphabet{\mathbfit}{OT1}{cmr}{bx}{it}
  \SetMathAlphabet\mathbfit{bold}{OT1}{cmr}{bx}{it}
  \DeclareMathAlphabet{\mathbfss}{OT1}{cmss}{bx}{n}
  \SetMathAlphabet\mathbfss{bold}{OT1}{cmss}{bx}{n}
  \ifAMStwofonts
    \ifCUPmtlplainloaded \else
      \DeclareSymbolFont{UPM}{U}{eur}{m}{n}
      \SetSymbolFont{UPM}{bold}{U}{eur}{b}{n}
      \DeclareSymbolFont{AMSa}{U}{msa}{m}{n}
      \DeclareMathSymbol{\upi}{0}{UPM}{"19}
      \DeclareMathSymbol{\umu}{0}{UPM}{"16}
      \DeclareMathSymbol{\upartial}{0}{UPM}{"40}
      \DeclareMathSymbol{\leqslant}{3}{AMSa}{"36}
      \DeclareMathSymbol{\geqslant}{3}{AMSa}{"3E}

      \let\leq=\leqslant 
      \let\geq=\geqslant 
    \fi
  \fi
\fi 

\ifCUPmtlplainloaded \else
  \ifAMStwofonts \else 
    \def\upi{\pi}
    \def\umu{\mu}
    \def\upartial{\partial}
  \fi
\fi

\title[I. Observations and nuclear data]
 {Group, field and isolated early-type galaxies \\
  I. Observations and nuclear data}

\author[G. Denicol\'o \etal]
    {G. Denicol\'o,$^1$\thanks{Visitor at INAOE, Mexico.} 
    Roberto Terlevich,$^2$\thanks{Visiting Fellow, IoA, Cambridge. E-mail for contact: {\it rjt@inaoep.mx}  \ and  \ {\it eterlevi@inaoep.mx}}  
    Elena Terlevich,$^2$\footnotemark[2]
    Duncan A. Forbes,$^3$ \cr
    Alejandro Terlevich,$^4$
    Luis Carrasco$^2$ \\
$^1$Institute of Astronomy, Madingley Road, Cambridge, CB3 0HA, United Kingdom \\
$^2$Instituto Nacional de Astrof\'{\i}sica, \'Optica y Electr\'onica, 
Tonantzintla, Puebla, Mexico \\
$^3$Centre for Astrophysics \& Supercomputing, Swinburne University, Hawthorn, VIC 3122, Australia\\
$^4$School of Physics and Astronomy, University of Birmingham, Edgbaston, Birmingham, B15 2TT, United Kingdom}

\date{Accepted. NOvember 3rd 2004 Received; in original form 2004 July}

\pagerange{\pageref{firstpage}--\pageref{lastpage}}
\pubyear{2004}

\begin{document}

\maketitle

\label{firstpage}


\begin{abstract}

This is the first paper of a series on the investigation of stellar population properties and galaxy evolution of an observationally homogeneous sample of early-type galaxies in groups, field and isolated galaxies.		

Here we present high signal-to-noise long-slit spectroscopy of 86 
nearby elliptical and S0 galaxies. Eight of them are isolated, selected 
according to a rigorous criterion, which
guarantees a genuine low-density sub-sample.  The present survey has the advantage of covering a larger wavelength range than normally found in the literature, which includes \Oiii$\lambda$5007 and H$\alpha$, both lines important for emission correction. Among the 86 galaxies with S/N $\geq$ 15 (per resolution element, for r$_e$/8 central aperture), 57 have their H$\beta$-index corrected for emission, the average correction is 0.190 \AA\ in H$\beta$; 42 galaxies reveal \Oiii$\lambda$5007 emission, of which 16 also show obvious  H$\alpha$ emission. Most of the galaxies in the sample do not show obvious signs of disturbances nor tidal features in the morphologies, 
although 11 belong to the Arp catalogue of peculiar galaxies; only three 
of them (NGC~750, NGC~751, NGC~3226) seem to be strongly interacting. We present the measurement of 25 {\it central} line-strength indices calibrated to the Lick/IDS system. Kinematic information is obtained for the sample. We analyse the line-strength index {\it vs} velocity dispersion relations for our sample of mainly low density environment galaxies, and compare the slope of the relations with cluster galaxies from the literature. Our main findings are that the index-$\sigma_0$ relations presented for low-density regions are not significantly different from those of cluster E/S0s. The slope of the index-$\sigma_0$ relations does not seem to change for early-type galaxies of different environmental densities, but the scatter of the relations seems larger for group, field and isolated galaxies than for cluster galaxies.

\end{abstract}



\section{Introduction}

The standard paradigm for galaxy formation is one of hierarchical
clustering and subsequent merging to form progressively larger
galaxies (White \& Rees 1978).
All galaxies are located within dark matter halos, and
the properties of galaxies depend on the assembly history of
these dark matter halos and hence their environment. This process has 
been modelled
by semi-analytical methods (e.g. Baugh \etal 1998;
Kauffmann et al. 1999; Somerville \& Primack 1999) and by
hydrodynamical simulations (e.g Cen \& Ostriker 1992;
Pearce \etal 1999; Berlind \etal 2003). 
For example, Kauffmann \& Charlot (1998) predict that galaxies in low 
density environments are younger (in a luminosity weighted sense) on 
average than those in clusters as they have a more extended merger history. 
They also predict that massive ellipticals are younger than smaller
ones, again due to merger-induced star formation. Quantitative
predictions depend on the details of the gas physics and feedback
processes (e.g. Kay \etal 2002; Kauffmann \etal 2004).
High quality data already exists for the cluster ellipticals, thus {\it determining the ages and metallicities of field and group ellipticals is a key test of the HCM paradigm} (Gonz\'alez 1993; Bower \etal 1998; 
Trager \etal 2000a,b; Kuntschner 2000; Terlevich \& Forbes 2002;  Chiosi \& Carraro 2002; Kuntschner \etal 2002; Willis \etal 2002).

In an effort to disentangle the degeneracy of age and metallicity effects on the gross properties of stellar populations, Worthey (1994, 1997) developed a series of stellar population spectral synthesis models. He showed that certain predominantly age (i.e. H$\beta$, H$\gamma$) and metallicity (i.e. Fe, [MgFe]) sensitive spectral absorption features can break the degeneracy, and the luminosity weighted mean ages and metallicities can be determined. In a study of bright galaxies in the Fornax cluster, Kuntschner \& Davies (1998) showed that, as predicted by the high {\it z} studies, the cluster ellipticals are indeed uniformly old, and span a range in metallicities. However, earlier studies of field (and loose group) ellipticals (Gonz\'alez 1993; Trager 1997) seem to show just the opposite, displaying a uniform metallicity, but spanning a range in ages. This suggests the possibility of an {\it environmental dependence} of the colour-magnitude relation (CMR), or maybe two completely different mechanisms operating in cluster and field environments, conspiring to produce similar CMRs. {\it The universality of the CMR from cluster core to the field environments, and its interpretation as a metallicity sequence, is fundamental to its placing a limit on the formation epoch of cluster elliptical galaxies to z $>$ 1} (e.g. Bower, Lucey \& Ellis 1992). If there is in fact a conspiracy between age and metallicity producing a tight CMR, this limit could be significantly loosened. 

We have carried out a large and observationally homogeneous survey of local early-type galaxies in low density environments with the aim of determining nuclear parameters, kinematic, age and metallicity gradients. This will help to infer whether the central values represent a small localised starburst or whether the whole galaxy was involved. The spectra, when combined with existing photometry, will allow us to investigate scaling relations such as the fundamental plane, Mg-$\sigma$ and colour-magnitude relations for field galaxies, and directly compare how these relations agree with galaxy formation histories, from their mean ages and metallicities. The luminosity weighted mean ages and metallicities of these galaxies will also allow us to directly test fundamental predictions of hierarchical models for structure formation in the universe. The emission line data will allow the discrimination among the different ionization mechanisms and provide information about the kinematical status of the ionized gas. In this paper we present the data, central measures and comparisons with previous work.

The paper is divided as follows. In Section 2 we describe the sample selection, and Section 3 details the observations and data reduction procedures. Section 4 describes the transformation of spectral index measurements to the Lick/IDS system, an improved method for emission correction of the Balmer line indices, and the derivation of errors from repeated measurements.
We explain the determination of kinematical properties and errors in Section 5. Section 6 presents the finally corrected line-strength indices.  
In Section 7 we analyse the index {\it vs} central velocity dispersion relations. Our results are summarized in Section 8. Complementary information about the sample and fully corrected Lick indices are presented in Appendices A and B.


\section{Selection of galaxies}

We have selected early-type galaxies that lie in relatively low density environments, i.e., in loose groups (56), the general field (19) and truly isolated 
galaxies (8).
However, we also observed 3 Virgo cluster galaxies, to complete a total of 86 galaxies. 
The objects are distributed all over the sky but with strong concentrations around the galactic poles and away from the galactic plane.
Galaxies in groups come from the group catalogue of Garcia (1993).

Eight galaxies were classified as truly isolated by Reda \etal (2004). Below, we describe briefly how isolated galaxies were selected. By using the {\it Lyon-Meudon Extragalactic Data Archive} (LEDA) of $\sim$100,000 galaxies, the following criteria were applied:

\begin{itemize} 
\item Morphological type $T\leq-3$			
\item Virgo corrected recession velocity $V\leq9000$ km s$^{-1}$, 
\item Apparent Magnitude $B_T\leq14.0$. 
\end{itemize}

This produced 330 galaxies which could be considered as potential isolated galaxies. The galaxies were compared to the rest of the catalogue and accepted as being isolated if they had no neighbours which were: 

\begin{itemize} 
\item within 700 km s$^{-1}$ in recession velocity, 
\item within 0.67 Mpc in the plane of the sky (assuming $H_0=75$ 
km s$^{-1}$ Mpc$^{-1}$),
\item less than 2 magnitudes fainter in $B_T$. 
\end{itemize}

Lastly, in the process of determining isolated objects, all galaxies were checked visually on the {\it Digitised Sky Survey} (DSS) images to ensure no near projected galaxy. This produced a final list of 40 early-type isolated galaxies; in this work we have obtained spectra of 8 of them. 

\

By exclusion, non-cluster galaxies which are not associated with groups and also not identified as isolated galaxies are then assumed to lie in the general field.

The Tully (1987) local density parameter $\rho$ is available for about 3/4 of our sample. Values range from 0.08 (isolated) to 3.99 (a Virgo galaxy). The sample also shows a good overlap with the Lick/IDS system, with 59 galaxies in common (Trager \etal 1998). 



\section{Observations and data reduction}

\begin{figure*}
\vspace{22.5 cm}
\includegraphics{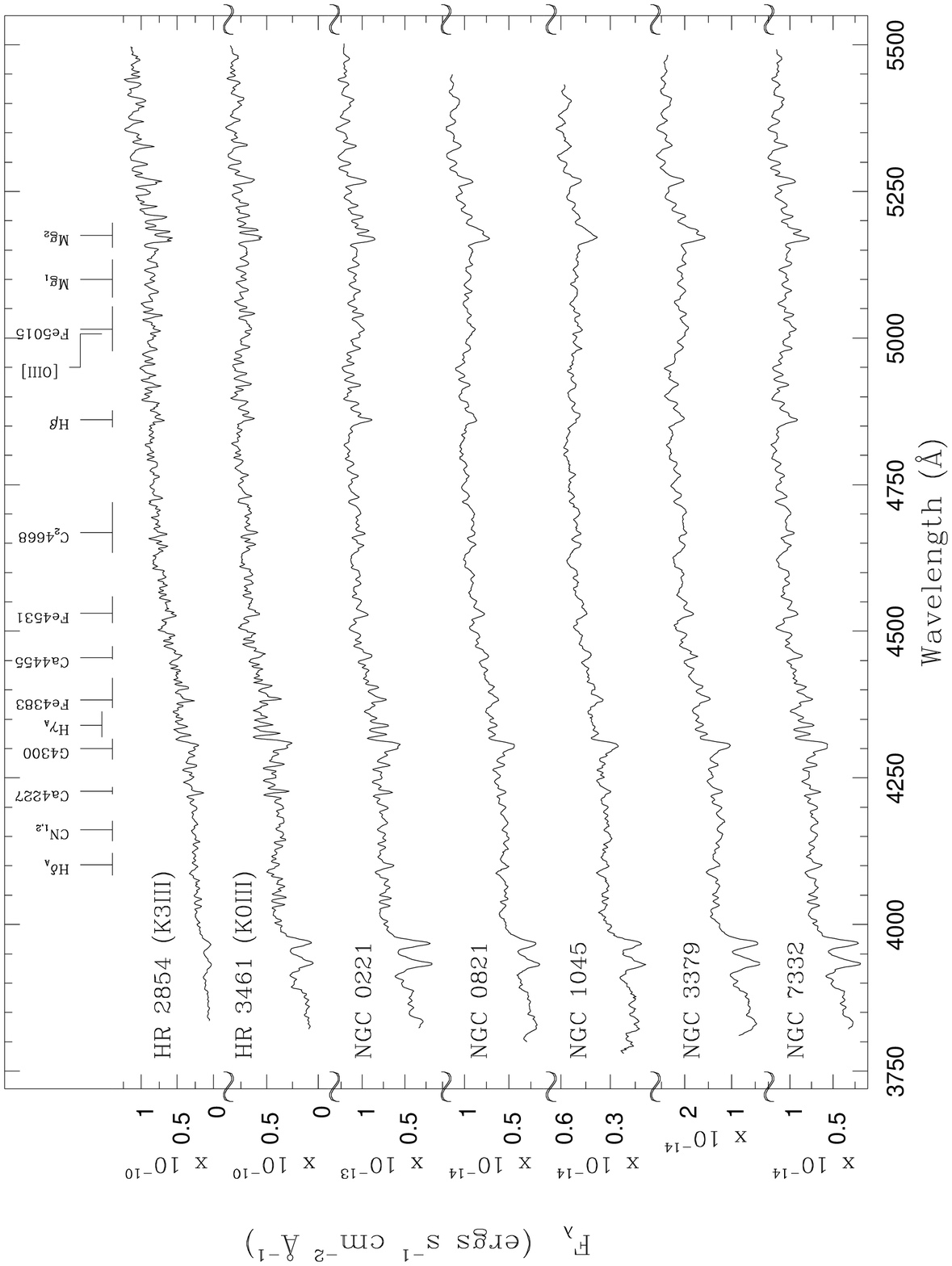}
\includegraphics{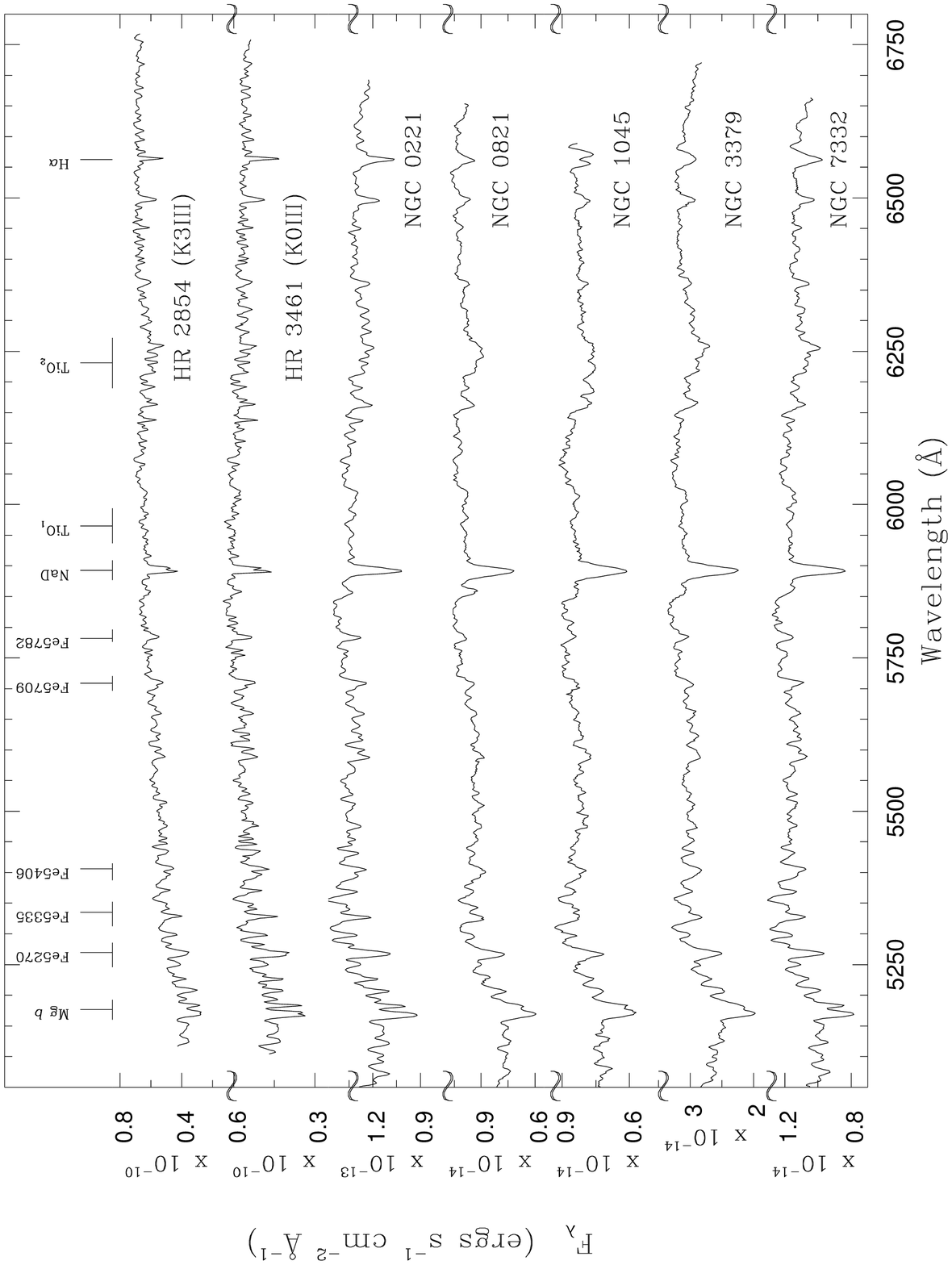}
\caption{Typical spectra of two stars and five galaxies. The top panel shows the spectra in the blue wavelength range, the bottom panel presents the red wavelength range. All spectra are flux calibrated using standard Oke (1990) stars. The galaxy spectra are redshift corrected and represent the nuclear r$_e$/8 extraction.}
\end{figure*}

Long-slit spectra have been obtained, during five different runs, accounting for a total of 30 nights, at the 2.12 m telescope of {\it Observatorio Astrof\'\i sico Guillermo Haro} (OAGH), in Cananea, Mexico. The telescope is equipped with a Boller \& Chivens spectrograph and a Tek 1024 pix$^2$ CCD camera. Parameters of the observations are reported in Table 1. We have obtained spectra in two wavelength ranges (blue and red), covering a total range from $\sim$ 3850 \AA\ to $\sim$ 6700 \AA. Two 300 lines mm$^{-1}$ grating blazed at 4650 \AA\ and 5900 \AA\ (for the blue and red wavelength ranges, respectively) yielded a dispersion of 65 \AA\ mm$^{-1}$ on the spectrograph CCD, i.e., an instrumental dispersion of $\sim$ 1.6 \AA/pixel. Typically, exposure times were between 300 to 1200 seconds per galaxy frame. Total exposure times were chosen to obtain S/N $>$ 40 per \AA\ at 5000\AA. The slit has been centred on the nucleus and oriented along the major axis for most of the galaxies. 
We observed the galaxies close to minimum zenithal distance in order to 
minimize differential refraction effects.
 Some neighbouring galaxies have been observed along the line connecting their nuclei. Instrumental resolution of $\sim$6 \AA\ (FWHM) was achieved with a slit width of 1.5 arcsec. Table 2 presents the detailed listing of the galaxy observations. The seeing was generally better than 2.0 arcsec. Additionally we observed 46 different standard stars (mainly K-giants) to be used as templates for velocity dispersion measurements, and for calibration of the line-strength indices to the Lick/IDS system (Gonz\'alez 1993). Six spectrophotometric standard stars were also taken to enable flux calibration. At least two different spectrophotometric standard stars were observed per night. Table 3 lists all observed stars with their spectral types (obtained from {\tt SIMBAD}, operated by CDS, Strasbourg). In addition, a bona fide elliptical galaxy, NGC~3379,  was observed in every run to serve as a link galaxy for the measurement of indices and velocity dispersion between the different observing runs. 

All data reduction was performed using packages in {\tt IRAF}. Each frame has been treated separately. Initial reduction of the CCD frames involved overscan correction, bias subtraction, the removal of pixel-to-pixel sensitivity variations (using dome flat exposures), and correction for uneven illumination along the slit (using twilight sky exposures). Careful wavelength calibration was performed using He-Ar or Fe-Ar lamp exposures. For the blue wavelength range observations ($\sim$3850-5450 [\AA]), He-Ar lamps were taken before and after each exposure to achieve an accurate wavelength calibration. For the red wavelength range ($\sim$5100-6700 [\AA]), we followed the same procedure using Fe-Ar lamps. A linear fit between pixel number and wavelength for $\sim$30 arc lines gave an rms calibration error $<$0.1 \AA\ in most cases. The galaxy frames were then corrected for atmospheric extinction, and afterwards, the continuum shape of the spectra was corrected to a relative flux scale with the help of the spectrophotometric standard stars. 
The cosmic rays were cleaned with the task {\tt crmedian}, which detects cosmic rays from pixels deviating by a 3-sigma level from the median at each pixel. 
Sky subtraction was then performed in each frame and finally, we used the task {\tt apall} to extract 1d-spectra. We have extracted the central r$_e$/8 (r$_e$ is the effective radius\footnote[2]{{\tt Effective radii are from Trager \etal 2000b and RC3.}}) from all galaxy frames, and these extractions will be used as our standard nuclear observations. A sample of seven typical spectra is shown in Figure 1.

The signal-to-noise ratio (S/N) per resolution element (resolution element = 6 \AA\ FWHM) of individual 1-D galaxy extractions ranges from 9 to about 100. The signal-to-noise was estimated assuming Poisson statistics,  $\sqrt{N}$, where N is the number of photon counts. The gain of the CCD, as presented in Table 1, is 1.85 e$^-$/ADU, and therefore the mean number of counts in the spectra was multiplied by the gain to obtain the photon statistics. 
The S/N measurement was performed in regions of $\sim$ 120 \AA\ encompassing the continua of Mg$_2$ index or the continua of TiO$_2$ index, for the blue or red spectra, respectively.

 We have only performed equivalent width measurements in spectra with S/N $\geq$ 15 per resolution element. 
We opted to work with the galaxy frames separately to avoid any possible wavelength mismatch when combining the frames. This decision offers statistical advantages as we are working with $\sim$ 700 spectral frames of S/N $\geq$ 15 (per resolution element) for a sample consisting of a total of 86 different galaxies (total of 8 frames/galaxy on average, where 4 frames are in the red wavelength range and 4 frames in the blue spectral range, in general). The line-strength index measurements were therefore performed in all spectra separately and later averaged for each galaxy. 

The median S/N per frame is $\sim$ 40 per resolution element (Figure 2). The effective S/N per galaxy can then be estimated as, approximately, the square-root of the number of frames per galaxy times the median S/N per frame, i.e., effective S/N per galaxy in the blue {\it or} red spectral range (average of 4 frames per galaxy {\it per spectral range}) $\approx$ $\sqrt{4} \times$ 40 $\simeq$ 80 per resolution element.  

Finally, we refer to the data in this paper as the OAGH sample.

\begin{table}
\centering
\begin{minipage}{200mm}
\caption{Instrumental set-up.}
Telescope \ \ \ \ \ \ \ \ \ \ \ \ \ \ \ \ \ \ \ \ \ \ \ OAGH 2.12 m, Cananea, Mexico\\
\begin{tabular}{@{}lll@{}}
\hline
\multicolumn{1}{l}{} &  
\multicolumn{1}{l}{Blue}  &
\multicolumn{1}{l}{Red} \\ 
\hline
Spectral range [\AA]         & 3850-5450             &  5100-6700           \\
                             &                       &                      \\
Date of observations         &{\tiny 2000 Mar 25-29}      & {\tiny 2000 Mar 30-31}     \\
                             &{\tiny 2000 Apr 01-02}      & {\tiny 2000 Apr 03}        \\
                             &{\tiny 2000 Oct 20-26}    & {\tiny 2001 Mar 27-29}     \\
                             &{\tiny 2001 Oct 13,18-19 } & {\tiny 2001 Oct 14-17}   \\
                             &{\tiny 2002 Apr 08}         & {\tiny 2002 Apr 06-07}     \\  
                             &                       &                      \\
Spectrograph                 & B \& C                & B \& C    \\
Detector                     & CCD Tek 1024          & CCD Tek 1024         \\ 
Gain [$e^-$/ADU]             & 1.85                  &  1.85                \\
Read-out-noise [$e^-$ rms]   & 3.7                   &  3.7                 \\
Pixel size [$\mu m^2$]       & 24                    &  24                  \\
Spatial scale [$''$/pixel]   & 0.44                  &  0.44                \\
Slit length [$'$]            & 3                     &  3                   \\
Slit width [$''$]            & 1.5                   &  1.5                 \\
Grating [lines/mm]           & 300                   & 300                  \\
Dispersion [\AA/pixel]       & $\sim$1.6             & $\sim$1.6           \\
Resolution [\AA]             & $\sim$6               & $\sim$6              \\
Seeing [arcsec]              & 1.5$-$2               & 1.5$-$2                \\
\hline
\end{tabular}
\end{minipage}
\end{table}

\begin{table*}
\centering
\begin{minipage}{200mm}
\caption{Log of observations: galaxies.}
\begin{tabular}{@{}llrrrr|cllrrrr@{}}
\hline
\multicolumn{1}{l}{Galaxy} &  
\multicolumn{1}{l}{Classification}  &
\multicolumn{1}{c}{B$_T$} &
\multicolumn{2}{c}{Exp. [sec]} &
\multicolumn{1}{r}{PA} &
\multicolumn{1}{c}\vline &
\multicolumn{1}{l}{Galaxy} &  
\multicolumn{1}{l}{Classification}  &
\multicolumn{1}{c}{B$_T$} &
\multicolumn{2}{c}{Exp. [sec]} &
\multicolumn{1}{r}{PA} \\
\multicolumn{1}{l}{} &  
\multicolumn{1}{l}{}  &
\multicolumn{1}{c}{[mag]} &
\multicolumn{1}{c}{Blue} &
\multicolumn{1}{r}{Red} &
\multicolumn{1}{r}{[$^o$]} &
\multicolumn{1}{c}\vline &
\multicolumn{1}{l}{} &  
\multicolumn{1}{l}{}  &
\multicolumn{1}{c}{[mag]} &
\multicolumn{1}{c}{Blue} &
\multicolumn{1}{r}{Red} &
\multicolumn{1}{r}{[$^o$]} \\ 
\hline
ESO 462-G015    & E3             & 12.95 &  3600 &  6300 & 166 & \vline & NGC 3599        & SA0            & 12.82 &  9600 &  3600 &  20 \\
MCG -01-02-018  & (R')SA(r)0      & 14.23 &  3600 &  3600 & 170 & \vline & NGC 3607        & SA0            & 10.82 &  2700 &  2700 & 120 \\
NGC 0016        & SAB0           & 13.00 &  3600 &  2700 &  20 & \vline & NGC 3608        & E2             & 11.70 &  3600 &  2700 &  75 \\
NGC 0221        & cE2            &  9.03 &  3600 &  2700 &  90 & \vline & NGC 3610        & E5             & 11.70 &  3600 &  3600 & 130 \\
NGC 0315        & E+ \tiny LINER-Sy1& 12.20 &  5400 &  2700 &  55 & \vline & NGC 3613        & E6             & 11.82 &  3600 &  1800 & 102 \\
NGC 0474        & SA(s)0         & 12.37 &  4800 &  2700 &  90 & \vline & NGC 3636        & E0             & 12.82 &  4800 &  2400 &  90 \\ 
NGC 0584        & E4             & 11.44 &  3600 &  2700 &  55 & \vline & NGC 3640        & E3             & 11.36 &  3600 &  2400 & 100 \\
NGC 0720        & E5             & 11.16 &  2700 &  2700 & 140 & \vline & NGC 3665        & SA(s)0         & 11.77 &  3600 &  1800 & 130 \\
NGC 0750        & E pec          & 12.89 &  4800 &  3600 & 166 & \vline & NGC 3923        & E4-5           & 10.80 &  1200 &  2700 &  45 \\
NGC 0751        & E pec          & 13.50 &  4800 &  3600 & 166 & \vline &                 &                &       &  1200 &       &  90 \\
NGC 0777        & E1             & 12.49 &  3600 &  4800 & 155 & \vline & NGC 3941        & SB(s)0 \tiny Sy2& 11.25 &  3600 &  1800 &  90 \\
NGC 0821        & E6             & 11.67 &  2700 &  2700 &  32 & \vline & NGC 4125        & E6 pec \tiny LINER& 10.65 &  4800 &  1800 &  80 \\ 
NGC 0890        & SAB(r)0        & 12.20 &  3600 &  2700 & 140 & \vline & NGC 4261        & E2-3 \tiny  LINER& 11.41 &  3600 &   900 &  90 \\
NGC 1045        & SA0 pec        & 13.52 &  3600 &  2700 &  40 & \vline & NGC 4365        & E3             & 10.52 &  3600 &  2700 &  40 \\
NGC 1052        & E4 \tiny LINER-Sy2& 12.08 &  2700 &  2700 & 122 & \vline & NGC 4374        & E1 \tiny LINER & 10.09 &  1800 &   900 &  90 \\
NGC 1132        & E              & 13.25 &  3600 &  6300 & 140 & \vline & NGC 4494        & E1-2           & 10.71 &  3600 &  1800 & 180 \\
NGC 1407        & E0             & 10.70 &  2700 &  1800 & 180 & \vline &                 &                &       &       &  1800 &  90 \\
NGC 1453        & E2-3           & 12.77&  3600 &  2700 &  45 & \vline & NGC 4550        & SB0            & 12.56 &  3600 &  1800 &  90 \\
NGC 1600        & E3             & 11.93 &  3600 &  2700 &  15 & \vline &                 &                &       &  1200 &       &  35 \\
NGC 1700        & E4             & 12.20 &  2400 &  6300 &  90 & \vline & NGC 4754        & SB(r)0         & 11.52 &  3600 &  1800 & 115 \\
NGC 1726        & SA(s)0         & 12.66 &  4500 &  5100 & 180 & \vline & NGC 5322        & E3-4 \tiny LINER& 11.14 &  4800 &  3600 &  95 \\
NGC 2128        & S0             & 13.60 &  8400 &  7200 & 140 & \vline & NGC 5353        & S0             & 11.96 &  4800 &  4800 & 180 \\
NGC 2300        & SA0            & 12.07 &  9900 &  9900 & 180 & \vline & NGC 5354        & SA0 \tiny LINER & 12.33 &  4800 &  3600 &  90 \\   
NGC 2418        & E              & 13.16 &  7500 &  6600 &  50 & \vline & NGC 5363        & I0 \tiny LINER  & 11.05 &  5400 &  2700 & 135 \\ 
                &                &       &  2700 &       &  35 & \vline & NGC 5444        & E+             & 12.78 &  4800 &  2400 &  67 \\
NGC 2513        & E              & 12.59 &  8400 &  3600 &  90 & \vline & NGC 5557        & E1             & 11.92 &  3600 &  1800 &  90 \\
NGC 2549        & SA(r)          & 12.19 &  3600 &  1800 &  90 & \vline & NGC 5576        & E3             & 11.85 &  3600 &  1800 &  90 \\
NGC 2768        & S0 \tiny LINER & 10.84 &  3600 &  1800 & 180 & \vline & NGC 5638        & E1             & 12.14 &  3600 &  1800 & 170 \\
                &                &       &  2700 &       &  90 & \vline & NGC 5812        & E0             & 12.19 &  4800 &  2400 &  90 \\
NGC 2872        & E2             & 12.86 &  4800 &  2400 & 110 & \vline & NGC 5813        & E1-2           & 11.45 &  7200 &  5400 & 145 \\
NGC 2911        & SA0 pec \tiny Sy& 12.50 &  6000 &  9300 &  50 & \vline & NGC 5831        & E3             & 12.45 &  6000 &  2400 &  90 \\
NGC 2974        & E4             & 12.30 &  3600 &  2400 & 135 & \vline & NGC 5845        & E:             & 13.50 &  8400 &  2400 & 120 \\
NGC 3091        & E3             & 12.13 &  2700 &  1800 &  90 & \vline & NGC 5846        & E0-1 \tiny LINER-HII& 11.05 &  7200 &  4500 & 180 \\
NGC 3098        & S0             & 12.89 &  6000 &  3600 & 180 & \vline &                 &                &       &       &  1800 &  90 \\
NGC 3115        & S0             &  9.87 &  2700 &  2280 & 135 & \vline & NGC 5846A      & cE2-3          & 14.10 &  7200 &  4500 & 180 \\
NGC 3139        & S0 pec         & 15.00 &  2700 &  1800 &  90 & \vline & NGC 5854        & SB(s)0         & 12.71 &  4800 &  3600 & 155 \\
NGC 3156        & S0             & 13.07 &  4800 &  2700 &  40 & \vline & NGC 5864        & SB(s)0         & 12.77 &  3600 &  2400 & 155 \\
NGC 3193        & E2             & 11.83 &  3600 &  2700 &  90 & \vline & NGC 5869        & S0:            & 12.91 &  2700 &  1800 & 120 \\
NGC 3226        & E2 pec \tiny LINER& 12.30 &  4800 &  2400 &  90 & \vline & NGC 5982        & E3             & 12.04 &  3600 &  1800 & 114 \\
NGC 3245        & SA(r)0 \tiny HII-LINER&11.70 &  3600 &  1800 &  90 & \vline & NGC 6172        & E+             & 13.83 &  8400 &  9900 &  80 \\
NGC 3377        & E5-6           & 11.24 &  3600 &  1800 & 135 & \vline & NGC 6411        & E              & 12.79 &  2700 &  7200 &  60 \\
NGC 3379        & E1 \tiny LINER & 10.24 &  8100 &  3600 &  90 & \vline & NGC 7302        & SA(s)0         & 13.58 &  4800 &  6300 & 100 \\
                &                &       &  3600 &  1800 &  70 & \vline & NGC 7332        & S0 pec         & 12.02 &  6300 &  5400 & 155 \\ 
NGC 3384        & SB0            & 10.85 &  2700 &  1800 &  53 & \vline & NGC 7454        & E4             & 12.78 &  3600 &  2700 & 148 \\
NGC 3412        & SB(s)0         & 11.45 &  3600 &  1800 &  45 & \vline & NGC 7585        & (R')SA(s)0 pec  & 12.50 &  3600 &  2700 & 115 \\
NGC 3414        & S0 pec         & 11.96 &  4800 &  3600 & 110 & \vline & NGC 7619        & E              & 12.10 &  7200 &  5400 &  30 \\
                &                &       &  2700 &       &  25 & \vline & NGC 7626        & E pec          & 12.16 &  7200 &  5400 &  90 \\
\hline
\end{tabular}

Classification and apparent magnitude (B$_T$) information from NED (2003).

\end{minipage}
\end{table*}

\begin{table*}
\centering
\begin{minipage}{140mm}
\caption{Log of observations: stars.}
\begin{tabular}{@{}lll|clll@{}}
\hline
\multicolumn{1}{l}{Star} &  
\multicolumn{1}{l}{Type}  &
\multicolumn{1}{l}{Comment} &
\multicolumn{1}{c}\vline &
\multicolumn{1}{l}{Star} &  
\multicolumn{1}{l}{Type}  &
\multicolumn{1}{l}{Comment} \\
\hline
HR 0166     & K0V              & Lick/IDS std & \vline & HR 6770     & G8III            & Lick/IDS std \\ 
HR 1015     & K3III            & Lick/IDS std & \vline & HR 6806     & K2V variable     & Lick/IDS std \\
HR 1907     & K0IIIb           & Lick/IDS std & \vline & HR 6817     & K1III            & Lick/IDS std \\
HR 2429     & K1III variable   & Lick/IDS std & \vline & HR 6872     & K2III variable   & Lick/IDS std \\
HR 2459     & K5III            & Lick/IDS std & \vline & HR 7149     & K2III            & Lick/IDS std \\
HR 2574     & K4III            & Lick/IDS std & \vline & HR 7176     & K1III            & Lick/IDS std \\
HR 2697     & K2III variable   & Lick/IDS std & \vline & HR 7185     & B5IV             &              \\
HR 2854     & K3III            & Lick/IDS std & \vline & HR 7317     & K4III            & Lick/IDS std \\
HR 2970     & G9III            & Lick/IDS std & \vline & HR 7480     & A3IV             &              \\
HR 3145     & K2III            & Lick/IDS std & \vline & HR 7576     & K3III            & Lick/IDS std \\
HR 3461     & K0III            & Lick/IDS std & \vline & HR 7596     & A0III            &              \\
HR 3845     & K2.5III variable & Lick/IDS std & \vline & HR 7914     & G5V              & Lick/IDS std \\
HR 3905     & K2III            & Lick/IDS std & \vline & HR 7957     & K0IV             & Lick/IDS std \\
HR 5370     & K3III variable   & Lick/IDS std & \vline & HR 8165     & K1III            & Lick/IDS std \\
HR 5480     & G7III variable   & Lick/IDS std & \vline & HR 8430     & F5V              & Lick/IDS std \\
HR 5744     & K2III variable   & Lick/IDS std & \vline & HR 8665     & F7V              & Lick/IDS std \\
HR 5854     & K2IIIb           & Lick/IDS std & \vline & HR 8772     & F8V              & Lick/IDS std \\
HR 5888     & G8III            & Lick/IDS std & \vline & HR 8832     & K3V              & Lick/IDS std \\
HR 5901     & K1IVa            & Lick/IDS std & \vline & HR 8841     & K0III            & Lick/IDS std \\
HR 5940     & K1IV             & Lick/IDS std & \vline & HR 8924     & K3III variable   & Lick/IDS std \\
HR 6014     & K1.5IV           & Lick/IDS std & \vline & HR 8969     & F7V              & Lick/IDS std \\
HR 6018     & K1III-IV variable& Lick/IDS std & \vline & BD+25 4655 & ?e...             & spec. std (Oke 1990) \\
HR 6136     & K4III            & Lick/IDS std & \vline & BD+33 2642 & B2IVp             & spec. std (Oke 1990) \\
HR 6159     & K4III variable   & Lick/IDS std & \vline & BD+75 325  & O5p variable      & spec. std (Oke 1990) \\
HR 6299     & K2III variable   & Lick/IDS std & \vline & Feige 34   & DA:               & spec. std (Oke 1990) \\
HR 6458     & G0V variable     & Lick/IDS std & \vline & G 191-B2B  & DAw...            & spec. std (Oke 1990) \\
HR 6710     & F2IV             & Lick/IDS std & \vline & Hz 44      & B2                & spec. std (Oke 1990) \\
\hline
\end{tabular}

Stellar type information from SIMBAD (2003).

\end{minipage}
\end{table*}

\begin{figure}
\vspace{7.5 cm}
\includegraphics{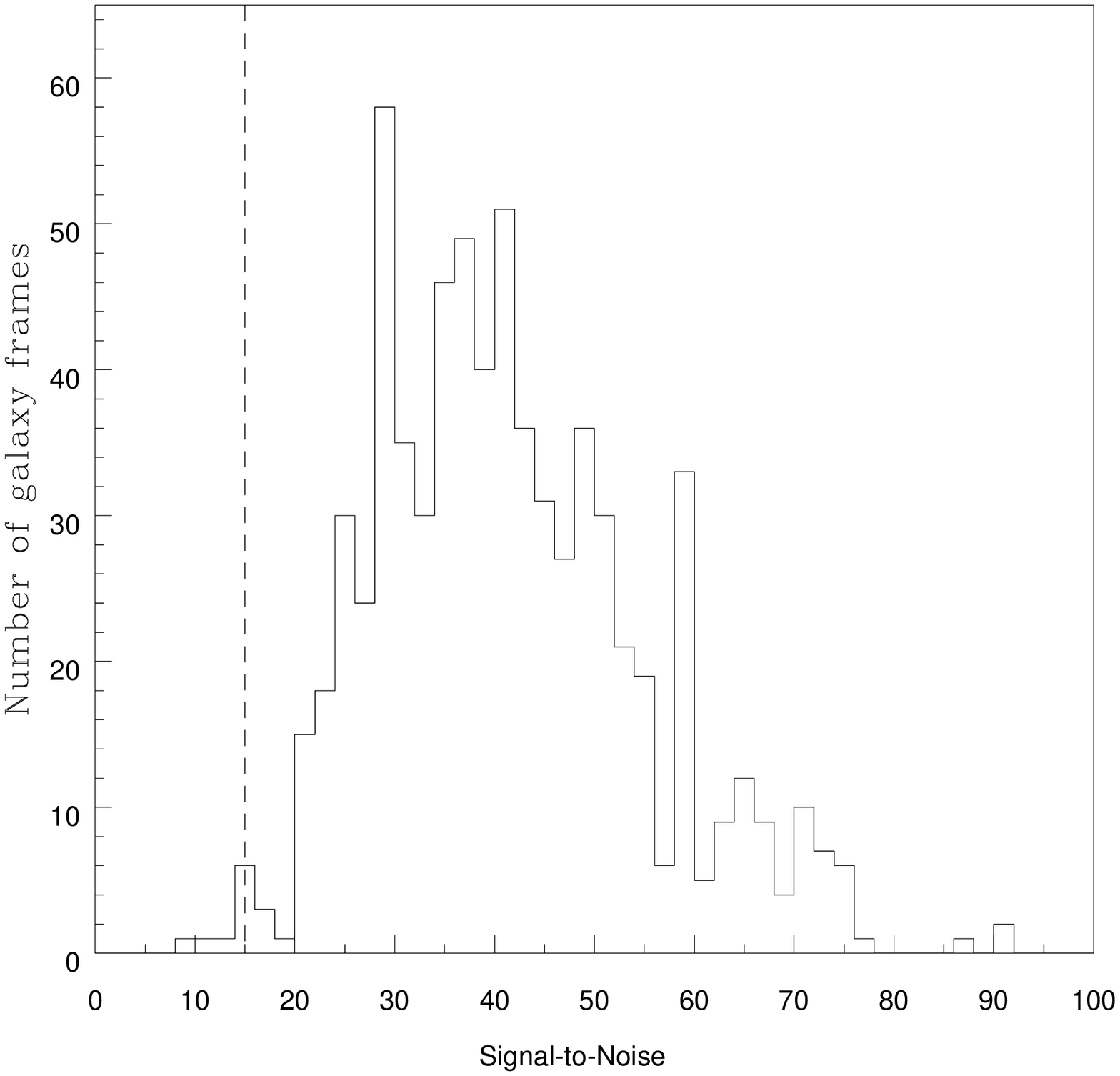}
\caption{Signal-to-noise ratio (per resolution element) for individual frames of our galaxy sample using r$_e$/8 central extractions. Index measurements were only performed in spectra with S/N $\geq$ 15 per resolution element (as shown by the dashed line).}
\end{figure}



\section{Lick Indices}

Absorption line indices in the visual as defined by the Lick group (e.g., H$\beta$, Mg$_2$, Fe5270, Fe5335, etc, Burstein \etal 1984; Faber \etal 1985) have proven to be a useful tool for the derivation of ages and metallicities of unresolved stellar populations.

Our spectra cover the 21 Lick indices defined in Faber \etal (1985) and Worthey \etal (1994), and the subsequent four Balmer indices (H$\delta_{A,F}$, H$\gamma_{A,F}$) first presented in Worthey \& Ottaviani (1997). The indices are calculated by the standard equations:

\begin{equation}
{EW_{\AA} = \int_{\lambda_1}^{\lambda_2} \left( 1- {F(\lambda) \over C(\lambda) } \right) d\lambda },
\end{equation}

\begin{equation}
{EW_{mag} = -2.5 \ log \left( {\int_{\lambda_1}^{\lambda_2} {F(\lambda) \over C(\lambda)}d\lambda } \over \lambda_2-\lambda_1 \right)},
\end{equation}

\noi{where}

\begin{equation}
{C(\lambda) = F_b {\lambda_r-\lambda \over {\lambda_r-\lambda_b}} +
 F_r {\lambda-\lambda_b \over {\lambda_r-\lambda_b}} },
\end{equation}

\noi{$\lambda_b$ and $\lambda_r$ are the mean wavelength in the blue and red pseudocontinuum intervals respectively.} We have adopted the spectral pseudocontinua and bandpasses of the 25 Lick/IDS indices defined in Worthey \etal (1994) and Worthey \& Ottaviani (1997).

The conversion between {\it suitable equivalent widths} (EW$_{\AA}$) and magnitude indices is given by

\begin{equation}
Index' = -2.5\ log\left(1-{Index \over \Delta\lambda}\right)
\end{equation}

\noi{where $\Delta\lambda$ is the width of the index bandpass.}

In this work we will use the index combination [MgFe] defined in Thomas, Maraston \& Bender (2003, TMB03. Note: in their paper this index is called [MgFe]$^\prime$, but we will adopt simply the notation [MgFe] for their same definition), and given by equation 5, which should be completely independent of $\alpha$/Fe variations, according to their models. We also use the index $<$Fe$>$ as the average given by equation 6 (defined by Kuntschner 1998):

\begin{equation}
[MgFe] = \sqrt{Mg{\it b} \cdot (0.72 \cdot Fe5270+0.28 \cdot Fe5335) }
\end{equation}

\begin{equation}
<Fe> = (Fe4383+Fe5270+Fe5335)/3
\end{equation}

Before measuring the indices, the galaxy spectra were convolved with gaussian curves in order to degrade their resolution to match the Lick/IDS data resolution. We have adopted the average values in the Lick/IDS resolution curve presented in Worthey \& Ottaviani (1997). The values we used are summarized in Table 4. We assume that the Lick/IDS resolution continues to degrade to the blue and red end, where we extrapolated the wavelength limits from Worthey \& Ottaviani (1997).

\begin{table}
\centering
\begin{minipage}{45mm}
\caption{Adopted resolutions to match Lick/IDS.}
\begin{tabular}{@{}lc@{}}
\hline
\multicolumn{1}{l}{Wavelength [\AA]} &
\multicolumn{1}{c}{FWHM [\AA]} \\
\hline 
$\sim$3850 \ - \ \ 4200  &    11.2   \\                               
\ \ 4200 \ - \ \ 4650    &     9.2   \\                             
\ \ 4651 \ - \ \ 5150    &     8.4   \\                          
\ \ 5151 \ - \ \ 5700    &     8.4   \\                         
\ \ 5701 \ - $\sim$6700  &     9.8   \\                 
\hline
\end{tabular}
\end{minipage}
\end{table}

\subsection{Index errors}

A central aspect of this work is that the determination of the errors in all measured and derived parameters is based on the analysis of the distribution of repeated observations.

We consider that the determination of errors through multiple observation is more reliable than estimates based on photon statistics of object and sky, detector noise, flat field errors, etc (e.g., Cardiel \etal 1998). This is because in the repeated observations all main sources of random error are included by default. Furthermore, no assumption about the error distribution function shape is necessary to determine its moments.

Care was taken in order to observe our sample of galaxies repeatedly and whenever possible in different nights or runs.
Only galaxies with a minimum of 3 independent observations are included in our sample.

We will consider throughout the paper that the {\it error of the mean} ($\sigma_{mean}$, resulting from the repeated measurement of indices in the various galaxy and stellar {\it frames}) is:

\begin{equation}
\sigma_{mean} = {{\it s} \over \sqrt{N}} 
\end{equation}
with,
\begin{equation}
{\it s} = \sqrt{{1 \over N-1} \sum^N_{i=1} (x_i-<x>)^2}
\end{equation}

{\noi where N is the number of frames per galaxy, {\it x$_i$} is the line-strength measurement in the spectrum {\it i}, $<$x$>$ is the mean of the measurements for a galaxy, and {\it s} is the standard deviation.}

We must emphasize that all error determinations originated from photon counts, spectral noise, etc, are mere error {\it estimates}; the {\it measurement} of an error is by definition obtained through repeated experiments. In this work we are presenting data which have the rare advantage of several repeated observations (measurements) per galaxy. On average, to each galaxy we have eight corresponding spectral frames. 
Therefore, for all the following analysis we will use $\sigma_{mean}$ as our error. 

Finally, for clarity purposes, all the errors presented in this paper (not only for the indices, but for all other measurements) are 1-sigma errors.

\subsection{Emission corrections}

Emission lines can be seen in many ellipticals at some level.
Spectroscopic surveys of early-type galaxies revealed that about 50-60\% of the galaxies show weak optical emission lines (Phillips \etal 1986; Caldwell 1984).
The origin of the gaseous component in ellipticals is not yet understood. It could be either of external origin, for example from a cooling flow or from 
a merger with a small gas rich galaxy; or it could be of internal origin, 
resulting from stellar mass loss. 

The emission-line spectra of giant elliptical galaxies are usually similar to those of LINERs (Low-Ionization Nuclear Emission Regions), where the ionization is provided by an energetic radiation from the nucleus, usually in the form of a power law. However, Filippenko \& Terlevich (1992), Binette \etal (1994) and Colina \& Arribas (1999) showed that post-AGB stars from the old stellar population of an early-type galaxy can also provide sufficient ionizing radiation to account for the observed H$\alpha$ luminosity and equivalent width.

More important for this study is that stellar absorption line-strength measurements can be affected if there is emission present in the galaxy, which fills the stellar absorptions.
In particular, the Balmer line indices used for age-dating stellar populations can be severely affected if there is emission, leading to wrong age estimates. 
Gonz\'alez (1993) proposed an empirical emission correction for H$\beta$ using the equivalent width of the emission line \Oiii$\lambda$5007 ($\Delta H\beta$ = 0.6 $\times$ \Oiii$\lambda$5007, as in Trager \etal 2000a). This turns out to be a very insecure correction because the emission spectra of \Hii regions are strong in \Hi recombination lines, but the strength of \Oiii$\lambda$4959 and \Oiii$\lambda$5007 lines can greatly differ (Osterbrock 1989; Carrasco \etal 1996). We note however, that on average Gonz\'alez corrections are shown to be appropriate (Trager \etal 2000a).

Unfortunately, most early-type galaxies that show signs of emission lines  
are discarded 
from further spectral analysis  in the literature, due to small 
wavelength coverage of the data (i.e. emission lines important to 
estimate corrections are not available) or the lack of a 
reliable emission correction technique. Our OAGH spectra 
instead, with the availability of the H$\alpha$ line information,
allow us to use a direct method for emission correction.

In our sample, of the 64 galaxies showing some emission, 29 
 have a final H$\beta$ correction $>$ 0.2 \AA. Twelve galaxies (NGC~2911 and 
NGC~3941 classified as Seyferts
and 10 LINERs) have been corrected using the H$\alpha$ 
emission correction.

  Table 5 presents the emission-line index definitions used here for H$\alpha$, \Nii$\lambda$6584 and \Oiii$\lambda$5007. These definitions work exactly as those of the Lick indices, with a central bandpass flanked by two pseudocontinua.

\subsubsection{Balmer lines}

Balmer line indices sense mainly the temperature of the turn off point from the main sequence and thus allow an age estimate of the integrated stellar population (e.g., Buzzoni \etal 1994). Note that the turn off temperature is also a function of metallicity and therefore the strength of the Balmer absorption is not only a function of age but also a function of metallicity (although to a smaller degree). In reality, the behaviour of Balmer lines at high metallicities seems to be uncertain. For example, Gregg (1994) found, analysing a sample of globular clusters, that the strength of the Balmer lines is only poorly correlated with metallicity above [Fe/H] = -0.7. There are some concerns that stars from other evolutionary phases (e.g., horizontal branch, HB) might contribute significantly to the Balmer absorption. A study of globular clusters by de Freitas Pacheco \& Barbuy (1995) showed that blue HB stars may give a substantial contribution to the H$\beta$ absorption (note that the higher order Balmer lines are more strongly affected by HB-contribution than H$\beta$; e.g., Peterson \etal 2003). Furthermore, they suggest that the HB morphology is correlated with the degree of central concentration in the globular cluster. Whether elliptical galaxies are subject to these effects is not yet known. Maraston \& Thomas (2000) have attempted to model the effect of HB morphology by calibrating the Balmer indices with globular cluster sample from Burstein \etal (1984), Covino \etal (1995) and Trager \etal (1998). In Maraston \etal (2003) it can be appreciated a good consistency between Balmer-line indices from models and observations. 
It is essential, however, to firstly correct the emission contamination in the Balmer indices before applying model predictions. Widely used in the literature, we are particularly interested in the H$\beta$ index.

Although the H$\beta$ index is potentially a good age indicator, is
the Balmer line more affected by the presence of nebular emission after H$\alpha$,
and if not corrected will cause a galaxy to appear older than it really is. 

\begin{table}
\centering
\begin{minipage}{90mm}
\caption{Emission index definitions.}
\vskip 0.2cm
\begin{tabular}{@{}lcr@{}}
\hline 
\multicolumn{1}{l}{Index} & 
\multicolumn{1}{c}{Central bandpass} & 
\multicolumn{1}{c}{Pseudocontinua} \\ 
\hline
\Oiii$\lambda$5007 &  4998.000--5015.000 &  4978.000--4998.000 \\
                   &                     &  5015.000--5030.000 \\ 
H$\alpha$          &  6558.000--6568.000 &  6372.000--6415.000 \\ 
                   &                     &  $\lambda_{red}$ = 6620.000  \\
\Nii$\lambda$6584  &  6574.000--6594.000 &  6372.000--6415.000 \\   
                   &                     &  $\lambda_{red}$ = 6620.000 \\    
\hline
\end{tabular}
\vskip 0.3cm
{\footnotesize{Note: We have assumed that the averaged flux calculated in the blue
pseudocontinua for H$\alpha$ and \Nii is equal to the averaged flux in the red continua, because in general the spectra did not have enough wavelength coverage to measure the red pseudocontinua for H$\alpha$ and \Nii. The central wavelength in the red used for the computation of equivalent widths was 6620 \AA.}}
\end{minipage}
\end{table}

As our wavelength range covers the H$\alpha$ line, we have corrected
H$\beta$ using EC(H$\beta$) (which stands for Emission Correction of the H$\beta$ index):

\begin{equation}
EC(H\beta) = {1\over
y}\cdot{C_{\alpha}\over{C_{\beta}}}\cdot[EW(H\alpha)_{tot}-EW(H\alpha)_{abs}]
\end{equation}

\noi{where}

\begin{equation}
C_{\alpha} = {Cont(H\alpha)\over{\Delta\lambda_{\alpha}}}  \ \ \  and \ \
\  C_{\beta} = {Cont(H\beta)\over{\Delta\lambda_{\beta}}}.
\end{equation}

\

\noi{Eq. 9 was derived assuming the following simplified definition
for equivalent width:}

\begin{equation}
EW(\lambda)_{tot} =
\left(1-{F(\lambda)_{tot}\over{Cont(\lambda)}}\right)\ \Delta\lambda,
\end{equation}

\noi{where}

\begin{equation}
F(\lambda)_{tot} = F(\lambda)_{em} + F(\lambda)_{abs}
\end{equation}

\noi{and}

\begin{equation}
F(H\alpha)_{em} = y \times F(H\beta)_{em}.
\end{equation}

\noindent{Cont(H$\alpha$) and Cont(H$\beta$) are the continua calculated for H$\alpha$
and H$\beta$ indices respectively. We assumed $y = 2.85$ for most
galaxies (Osterbrock 1989, Case B of recombination, low density limit at electron temperature of 
10$^4$ K).
In the equations above, the subscript
{\it tot} stands for total observed values, {\it em} and {\it abs} are
emission and absorption, respectively. $EC(H\beta)$ is the emission
correction in \AA ngstroms for H$\beta$. {\it We have assumed
$EW(H\alpha)_{abs}$ = 1.03 $\pm$ 0.08 \AA}\footnote[2]{{\tt $\pm$ 0.08 represents the peak to peak range of measurements. Ideally this value should be obtained also from spectral synthesis simulations but there is no simulation yet published with resolution high enough at H$\alpha$ to do this estimate.}}, derived from EW(H$\alpha$)
measurements in our sample of mainly K stars. Note that the uncertainty in $EW(H\alpha)_{abs}$ takes into account how much a reasonable range of values for $EW(H\alpha)$ in absorption would change the correction.  

We have corrected the Balmer lines using the term $EC(H\beta)$. The actual correction
has been applied to all but 9 galaxies, for which the correction calculated is very small and 
consistent with zero, and the errors 
of the correction are larger than the correction itself. 

In Figure 3 we show how much the emission correction from eq. 9 would change if we assumed $EW(H\alpha)_{abs}$ = 0.95 \AA\ or $EW(H\alpha)_{abs}$ = 1.11 \AA. The difference in the emission correction for both 
$EW(H\alpha)_{abs}$ values is generally smaller than 0.1 \AA. But this difference is for an extreme variation of $EW(H\alpha)_{abs}$ and must be regarded as an upper limit.

In the expressions above, emission is considered to be negative and
absorption positive, therefore $\mid$$EC(H\beta)$$\mid$ should be subtracted from
$EW(H\beta)_{tot}$. In an attempt to compensate for reddening effects
in eq. 9, we have assumed a value of 3.0 for the ratio of
intensities F(H$\alpha$)$_{em}$/F(H$\beta$)$_{em}$ of a few dusty
galaxies: NGC~2872, NGC~3226, NGC~4125, NGC~4494, NGC~5812, NGC~5813
and NGC~5846 (e.g.~Forbes 1991,  Tran \etal 2001). The emission
correction to H$\beta$ using the H$\alpha$ line is presented in column
2 of Table 6.

Trager \etal (2000a) have corrected the H$\beta$ data for emission by subtracting 0.6 $\times$ \Oiii$\lambda$5007 emission to the H$\beta$
index. This empirical correction is believed to be valid only in a
statistical sense, and not accurate for individual galaxies. For
comparative purposes, we have measured the \Oiii index as defined in
Kuntschner \etal (2001) and derived the associated emission correction, which is
listed in column 3 of Table 6.

\begin{figure}
\vspace{7.5 cm}
\includegraphics{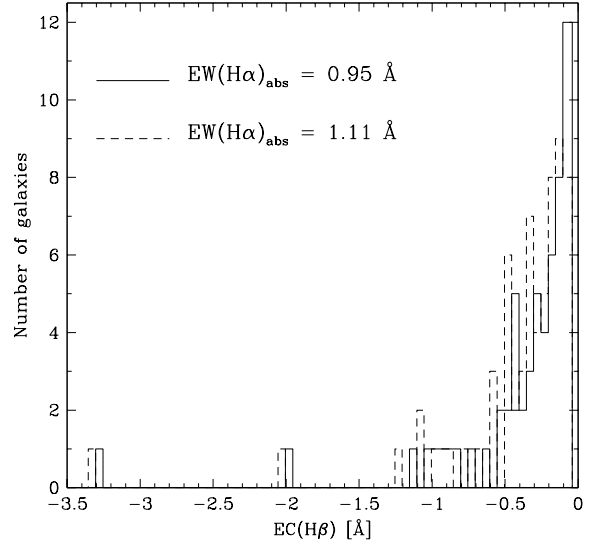}
\caption{ Histogram of emission correction EC(H$\beta$) assuming two different values for $EW(H\alpha)_{abs}$. When we take $EW(H\alpha)_{abs}$ = 0.95 \AA\ the emission correction is in general less than 0.1 \AA\ smaller than for $EW(H\alpha)_{abs}$ = 1.11 \AA. The average value for $EW(H\alpha)_{abs}$ used throughout the paper is 1.03 $\pm$ 0.08 \AA. }
\end{figure}

Comparing the different emission corrections, the H$\alpha$ and the empirical \Oiii$\lambda$5007 corrections,
 the difference is small for the majority of galaxies that show emission in the \Oiii$\lambda$5007 forbidden line (see Figure 4-c).
 We emphasize that it is very dangerous to rely on the \Oiii$\lambda$5007 alone because of
the uncertainties in the ionization parameter (i.e., one should not assume the same \Oiii/H$\beta$ ratio for all galaxies).
In principle, the H$\alpha$-based emission correction has less uncertainties
than using \Oiii$\lambda$5007.
Note that the errors for the \Oiii
correction in Table 6 are only the propagation of the error from the repeated  measurements of the emission line \Oiii$\lambda$5007: further errors are expected
since the \Oiii$\lambda$5007/H$\beta$ ratio may vary greatly from
galaxy to galaxy.

  Figure 4 shows the difference between the two emission correction methods for H$\beta$. Panels (a) and (b) demonstrate the effects of the emission correction in the H$\beta$ {\it vs} [MgFe] plane. Panel (c) compares the  H$\beta$ index derived from the \Oiii (i.e., H$\beta$+EC$_{\Oiii}$) and H$\alpha$-based (i.e., H$\beta$+EC$_{H\alpha}$) corrections. The H$\alpha$-based correction (EC$_{H\alpha}$) is larger than the \Oiii one for most cases, the difference can be up to $\sim$ 1.0 \AA.

Panel (d) shows the comparison between our measurements of \Oiii and the values in Trager \etal (2000a). The agreement is generally good if we consider the error bars. 
 Note however that we are using an index definition to measure \Oiii (Table 5) whereas Trager \etal (2000a, T00a) did not use the same definition. The data analysed in T00a is actually a compilation of the data of Gonz\'alez (1993), where only the emission profile of the \Oiii line was measured.

For the standard elliptical NGC~3379, the H$\beta$ line index was found to be 1.33 $\pm$ 0.09 \AA\ (no emission correction in this value) by Kuntschner (1998), whereas we measure H$\beta$ = 1.40 $\pm$ 0.06 \AA\ (raw value without emission correction). We have measured an emission correction of 0.10 $\pm$ 0.08 \AA\ using the H$\alpha$ method, and Kuntschner detected $\sim$ 0.12 \AA\ emission correction for NGC~3379 (using 0.6$\times$\Oiii$\lambda$5007).

The other Balmer-line indices, H$\gamma$ and H$\delta$, were also corrected using the same procedure as in eq. 9, but assuming $y_{\gamma}$ = 2.85/0.468 $\simeq$ 6.09 or y$_{\delta}$ = 2.85/0.259 $\simeq$ 11.00  for F(H$\alpha$)$_{em}$/F(H$\gamma$)$_{em}$ or F(H$\alpha$)$_{em}$/F(H$\delta$)$_{em}$ ratios, respectively (Osterbrock 1989; Case B of recombination, for electron density n$_e$ $<$ 10$^3$ and T$_{eff}$ = 10$^4$ K), and changing Cont(H$\beta$) for Cont(H$\gamma$) or Cont(H$\delta$) accordingly.

The galaxies NGC~1045 and NGC~1453, for which our spectra do not cover the H$\alpha$ line, have been emission corrected for H$\beta$ using the 0.6 $\times$ \Oiii$\lambda$5007 approximation. In the same way, H$\gamma_{A,F}$ and H$\delta_{A,F}$ for these galaxies have been corrected by the approximated scalings 0.36 $\times$ \Oiii and 0.22 $\times$ \Oiii, respectively, adopted in Kuntschner \etal (2002).

\begin{figure}
\vspace{21 cm}
\includegraphics{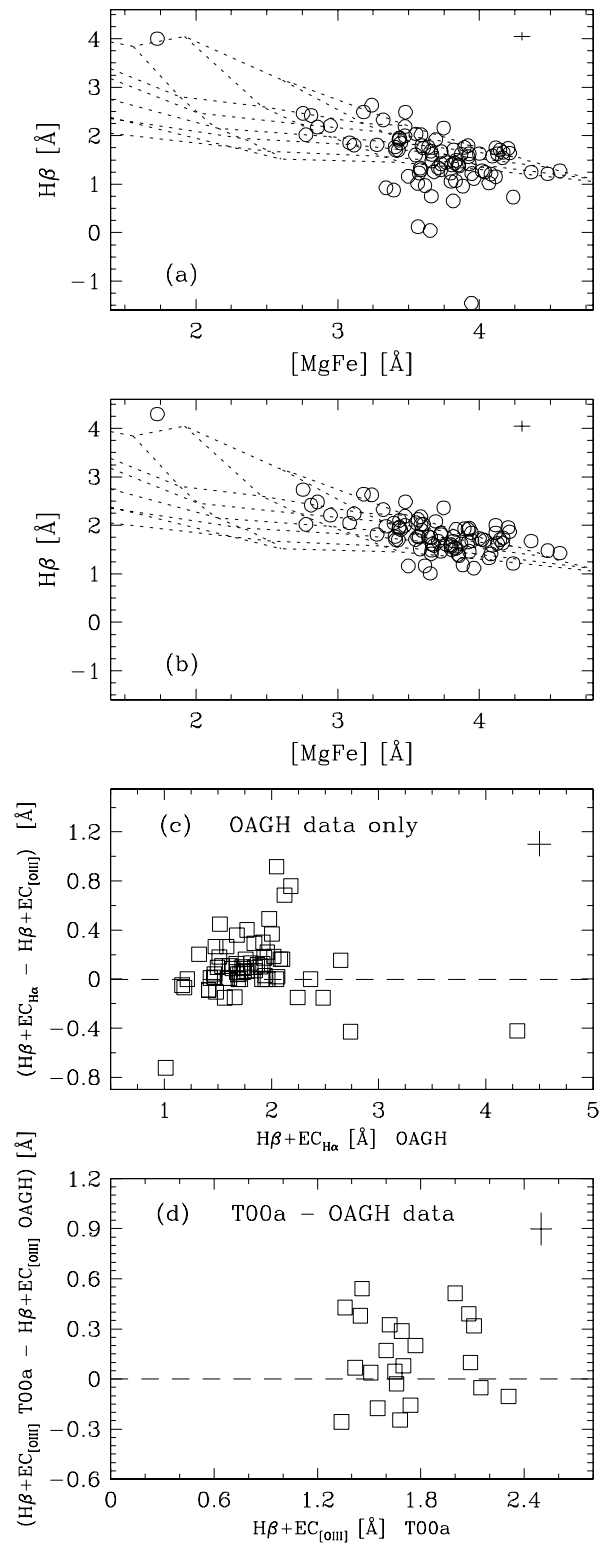}
\caption{Panels (a) and (b) demonstrate the effect of emission correction in the H$\beta$ {\it vs} [MgFe] plane: (a) no correction, (b) EC(H$\beta$) correction applied (Table 6). The dotted lines are single-stellar population models from Thomas \etal (2003) for a range of ages and metallicities. Panel (c) compares the  H$\beta$ index derived with the \Oiii and H$\alpha$-based corrections. Panel (d) presents the comparison between our measurements of \Oiii$\lambda$5007 and the values in Trager \etal (2000a, T00a). The error bars on the top right corner of the plots are averages of the error of the mean.}
\end{figure}

\subsubsection{Fe5015}

Emission can also occur in one of the side-bands of the Lick/IDS indices and raise the continuum which in turn gives larger index values.
As pointed out by Kuntschner \etal (2002), the Fe5015 index is affected by \Oiii$\lambda$5007 emission in its central bandpass, and by \Oiii$\lambda$4959 emission in its blue continuum bandpass. These authors have derived a correction to Fe5015 by artificially adding spectral emission and examining the effects on the Fe5015 index. We have used their result, where the index can be corrected by adding $+0.61(\pm0.01)\times\Oiii\lambda5007$ to the Fe5015 measurement. This procedure has large uncertainties and therefore Fe5015 measurements in ellipticals that show emission, either in H$\alpha$ or \Oiii, should be considered with care.

\begin{table*}
\scriptsize
\caption{Emission corrections in \AA ngstroms for the galaxies with detected emission lines in the central r$_e$/8 aperture extractions.}
\vspace{0.2 cm}
\centering
\begin{minipage}{160mm}
\begin{tabular}{@{}lcccccccc@{}}
\hline
\multicolumn{1}{l}{Galaxy} &
\multicolumn{1}{c}{EC(H$\beta$)} &
\multicolumn{1}{c}{0.6 $\times$ \Oiii$\lambda$5007} &
\multicolumn{1}{c}{EW(H$\alpha$)$_{tot}$} &
\multicolumn{1}{c}{EW(\Nii$\lambda$6584)} &
\multicolumn{1}{c}{EC(H$\gamma_A$)} &
\multicolumn{1}{c}{EC(H$\delta_A$)} \\
\multicolumn{1}{c}{(1)} &
\multicolumn{1}{c}{(2)} &
\multicolumn{1}{c}{(3)} &
\multicolumn{1}{c}{(4)} &
\multicolumn{1}{c}{(5)} &
\multicolumn{1}{c}{(6)} &
\multicolumn{1}{c}{(7)} \\
\hline      
NGC~0315  & -0.85 $\pm$  0.09 & -0.48 $\pm$  0.16 & -1.01 $\pm$  0.02 &              --    &  -0.71 $\pm$  0.11 &  -0.50 $\pm$  0.11 \\
NGC~0474  & -0.11 $\pm$  0.08 & -0.06 $\pm$  0.06 &  0.77 $\pm$  0.04 &              --    &             --     &               --     \\
NGC~0584  & -0.10 $\pm$  0.08 &            --     &  0.79 $\pm$  0.01 &  -0.29 $\pm$  0.02 &             --     &               --     \\
NGC~0720  & -0.09 $\pm$  0.08 &            --     &  0.82 $\pm$  0.02 &              --    &             --     &               --     \\
NGC~0821  & -0.13 $\pm$  0.08 &            --     &  0.71 $\pm$  0.05 &  -0.23 $\pm$  0.08 &             --     &               --     \\
NGC~1045  &    *              & -0.32 $\pm$  0.08 &        *          &         *          &           *        &              *       \\
NGC~1052  & -3.30 $\pm$  0.40 & -3.23 $\pm$  0.09 & -6.89 $\pm$  0.20 & -12.22 $\pm$  0.18 &  -2.75 $\pm$  0.52 &  -1.95 $\pm$  0.54 \\
NGC~1407  & -0.22 $\pm$  0.07 &            --     &  0.50 $\pm$  0.03 &  -0.13 $\pm$  0.01 &             --     &               --     \\
NGC~1453  &     *             & -0.48 $\pm$  0.08 &        *          &         *          &           *        &              *      \\
NGC~1600  & -0.23 $\pm$  0.07 & -0.11 $\pm$  0.10 &  0.49 $\pm$  0.04 &  -0.23 $\pm$  0.10 &             --     &               --     \\
NGC~1700  & -0.16 $\pm$  0.08 &            --     &  0.64 $\pm$  0.01 &  -0.52 $\pm$  0.01 &             --     &               --     \\
NGC~1726  & -0.33 $\pm$  0.09 & -0.21 $\pm$  0.12 &  0.23 $\pm$  0.15 &  -0.87 $\pm$  0.15 &             --     &               --     \\
NGC~2128  & -1.17 $\pm$  0.12 & -0.25 $\pm$  0.10 & -1.78 $\pm$  0.05 &  -3.88 $\pm$  0.03 &  -0.97 $\pm$  0.16 &  -0.69 $\pm$  0.16 \\
NGC~2300  & -0.07 $\pm$  0.09 &            --     &  0.88 $\pm$  0.05 &  -0.38 $\pm$  0.12 &             --     &               --     \\
NGC~2418  & -0.47 $\pm$  0.07 & -0.43 $\pm$  0.06 & -0.09 $\pm$  0.04 &  -1.36 $\pm$  0.04 &  -0.39 $\pm$  0.08 &  -0.28 $\pm$  0.08 \\
NGC~2513  & -0.09 $\pm$  0.09 &            --     &  0.82 $\pm$  0.07 &              --    &             --     &               --     \\
NGC~2549  & -0.20 $\pm$  0.13 & -0.20 $\pm$  0.13 &  1.26 $\pm$  0.01 &              --    &             --     &               --     \\
NGC~2768  & -0.55 $\pm$  0.07 & -0.54 $\pm$  0.11 & -0.29 $\pm$  0.05 &  -1.59 $\pm$  0.07 &  -0.46 $\pm$  0.09 &  -0.33 $\pm$  0.09 \\
NGC~2872  & -0.11 $\pm$  0.09 &            --     &  0.77 $\pm$  0.08 &  -0.16 $\pm$  0.08 &             --     &               --     \\
NGC~2911  & -2.00 $\pm$  0.23 & -1.32 $\pm$  0.14 & -3.77 $\pm$  0.13 &  -9.55 $\pm$  0.18 &  -1.66 $\pm$  0.29 &  -1.18 $\pm$  0.31\\
NGC~2974  & -0.66 $\pm$  0.08 & -0.75 $\pm$  0.09 & -0.56 $\pm$  0.02 &  -3.40 $\pm$  0.11 &  -0.55 $\pm$  0.09 &  -0.39 $\pm$  0.09\\
NGC~3091  & -0.15 $\pm$  0.08 & -0.14 $\pm$  0.08 &  0.67 $\pm$  0.06 &              --    &             --     &               --    \\
NGC~3098  & -0.19 $\pm$  0.11 & -0.19 $\pm$  0.11 &  1.35 $\pm$  0.01 &              --    &             --     &               --    \\
NGC~3115  & -0.10 $\pm$  0.08 &            --     &  0.80 $\pm$  0.03 &              --    &             --     &               --    \\
NGC~3139  & -0.03 $\pm$  0.09 &            --     &  0.95 $\pm$  0.05 &              --    &             --     &               --    \\
NGC~3156  & -0.29 $\pm$  0.09 & -0.71 $\pm$  0.07 &  0.33 $\pm$  0.13 &  -0.96 $\pm$  0.02 &             --     &               --    \\
NGC~3193  & -0.09 $\pm$  0.08 & -0.24 $\pm$  0.01 &  0.81 $\pm$  0.03 &              --    &             --     &               --    \\
NGC~3226  & -0.97 $\pm$  0.10 & -1.69 $\pm$  0.05 & -1.29 $\pm$  0.02 &  -2.86 $\pm$  0.13 &  -0.80 $\pm$  0.13 &  -0.57 $\pm$  0.13\\
NGC~3245  & -0.75 $\pm$  0.08 & -0.35 $\pm$  0.07 & -0.78 $\pm$  0.02 &  -1.64 $\pm$  0.12 &  -0.63 $\pm$  0.10 &  -0.45 $\pm$  0.10\\
NGC~3377  & -0.01 $\pm$  0.04 & -0.01 $\pm$  0.04 &  1.37 $\pm$  0.02 &              --    &             --     &               --    \\
NGC~3379  & -0.10 $\pm$  0.08 &            --     &  0.80 $\pm$  0.03 &              --    &             --     &               --    \\
NGC~3414  & -1.02 $\pm$  0.11 & -0.96 $\pm$  0.07 & -1.42 $\pm$  0.06 &  -3.42 $\pm$  0.08 &  -0.85 $\pm$  0.13 &  -0.60 $\pm$  0.14\\
NGC~3599  & -0.28 $\pm$  0.08 & -0.70 $\pm$  0.02 &  0.37 $\pm$  0.07 &  -1.60 $\pm$  0.13 &             --     &               --    \\
NGC~3607  & -0.43 $\pm$  0.07 & -0.13 $\pm$  0.01 &  0.01 $\pm$  0.01 &  -2.21 $\pm$  0.02 &             --     &               --    \\
NGC~3608  & -0.16 $\pm$  0.08 & -0.06 $\pm$  0.08 &  0.64 $\pm$  0.03 &  -0.19 $\pm$  0.04 &             --     &               --    \\
NGC~3613  & -0.06 $\pm$  0.09 &            --     &  0.88 $\pm$  0.06 &              --    &             --     &               --    \\
NGC~3636  & -0.06 $\pm$  0.05 & -0.06 $\pm$  0.05 &  1.14 $\pm$  0.11 &              --    &             --     &               --    \\
NGC~3640  & -0.02 $\pm$  0.09 &            --     &  0.98 $\pm$  0.04 &              --    &             --     &               --    \\
NGC~3665  & -0.88 $\pm$  0.09 & -0.13 $\pm$  0.01 & -1.09 $\pm$  0.02 &  -1.15 $\pm$  0.05 &  -0.74 $\pm$  0.11 &  -0.52 $\pm$  0.12\\
NGC~3923  & -0.13 $\pm$  0.08 &         --        &  0.73 $\pm$  0.02 &           --       &         --         &            --         \\
NGC~3941  & -0.28 $\pm$  0.07 & -0.09 $\pm$  0.11 &  0.37 $\pm$  0.02 &  -0.56 $\pm$  0.04 &         --         &            --         \\
NGC~4125  & -0.53 $\pm$  0.07 & -0.37 $\pm$  0.20 & -0.23 $\pm$  0.04 &  -2.64 $\pm$  0.10 &  -0.44 $\pm$  0.08 &  -0.31 $\pm$  0.09 \\
NGC~4261  & -0.21 $\pm$  0.08 & -0.29 $\pm$  0.05 &  0.54 $\pm$  0.08 &  -1.44 $\pm$  0.10 &         --         &            --         \\
NGC~4365  & -0.10 $\pm$  0.08 &         --        &  0.80 $\pm$  0.02 &           --       &         --         &            --         \\
NGC~4374  & -0.55 $\pm$  0.08 & -0.25 $\pm$  0.03 & -0.28 $\pm$  0.07 &  -1.63 $\pm$  0.04 &  -0.45 $\pm$  0.09 &  -0.32 $\pm$  0.09 \\
NGC~4550  & -0.44 $\pm$  0.08 & -0.59 $\pm$  0.09 & -0.02 $\pm$  0.11 &  -0.27 $\pm$  0.08 &  -0.36 $\pm$  0.09 &  -0.26 $\pm$  0.09 \\
NGC~4754  & -0.01 $\pm$  0.09 &         --        &  1.01 $\pm$  0.04 &           --       &         --         &            --         \\
NGC~5322  & -0.16 $\pm$  0.08 &         --        &  0.64 $\pm$  0.02 &  -0.50 $\pm$  0.02 &         --         &            --         \\
NGC~5353  & -0.43 $\pm$  0.07 & -0.07 $\pm$  0.06 &  0.01 $\pm$  0.04 &  -1.42 $\pm$  0.10 &         --         &            --         \\
NGC~5354  & -0.19 $\pm$  0.08 & -0.24 $\pm$  0.12 &  0.57 $\pm$  0.03 &  -0.14 $\pm$  0.08 &         --         &            --         \\
NGC~5363  & -1.05 $\pm$  0.11 & -0.56 $\pm$  0.05 & -1.50 $\pm$  0.06 &  -3.12 $\pm$  0.02 &  -0.88 $\pm$  0.14 &  -0.62 $\pm$  0.14 \\
NGC~5444  & -0.23 $\pm$  0.08 & -0.29 $\pm$  0.06 &  0.49 $\pm$  0.05 &  -0.42 $\pm$  0.10 &         --         &            --         \\
NGC~5813  & -0.33 $\pm$  0.07 & -0.06 $\pm$  0.06 &  0.25 $\pm$  0.05 &  -0.77 $\pm$  0.14 &         --         &            --         \\
NGC~5831  & -0.05 $\pm$  0.09 & -0.03 $\pm$  0.10 &  0.92 $\pm$  0.07 &           --       &         --         &            --         \\
NGC~5845  & -0.17 $\pm$  0.08 &        --         &  0.61 $\pm$  0.05 &  -0.31 $\pm$  0.05 &         --         &            --         \\
NGC~5846  & -0.45 $\pm$  0.07 &        --         & -0.05 $\pm$  0.03 &  -1.82 $\pm$  0.08 &  -0.37 $\pm$  0.08 &  -0.27 $\pm$  0.08 \\
NGC~5846A & -0.01 $\pm$  0.09 & -0.11 $\pm$  0.04 &  1.02 $\pm$  0.03 &           --       &         --         &            --         \\
NGC~5869  & -0.33 $\pm$  0.51 & -0.33 $\pm$  0.51 &  1.14 $\pm$  0.06 &           --       &         --         &            --         \\
NGC~5982  & -0.04 $\pm$  0.09 &        --         &  0.93 $\pm$  0.01 &           --       &         --         &            --         \\
NGC~6172  & -0.30 $\pm$  0.07 & -0.46 $\pm$  0.10 &  0.30 $\pm$  0.04 &  -1.74 $\pm$  0.08 &         --         &            --         \\
NGC~7302  & -0.03 $\pm$  0.09 &        --         &  0.96 $\pm$  0.08 &  -0.29 $\pm$  0.06 &         --         &            --         \\
NGC~7585  & -0.16 $\pm$  0.08 &        --         &  0.66 $\pm$  0.02 &  -1.19 $\pm$  0.09 &         --         &            --         \\
NGC~7619  & -0.27 $\pm$  0.07 &        --         &  0.40 $\pm$  0.03 &  -0.33 $\pm$  0.05 &         --         &            --         \\
NGC~7626  & -0.30 $\pm$  0.07 & -0.10 $\pm$  0.07 &  0.31 $\pm$  0.03 &        --          &         --         &            --       \\
\hline
\end{tabular}
\vspace{0.2 cm}

\footnotesize{Notes: EC($\lambda$) was derived using EW(H$\alpha$)$_{abs}$ = 1.03 \AA\ (see eq. 9). The * symbol is used when H$\alpha$ and/or \Nii fell outside the wavelength limits of the spectra and could not be measured. The --- symbol denotes no emission component detected (if EC(H$\beta$) shows -- it is because no emission was detected in H$\alpha$). The negative signs denote emission, therefore, the EC($\lambda$) corrections must be subtracted from the effective index measurement, e.g., H$\beta_{correct}$ = H$\beta$--EC(H$\beta$). Columns 4 and 5 are EW(H$\alpha$) and EW(\Nii) indices, measured in \AA ngstroms, and defined in Table 5.}
\end{minipage}
\end{table*}
\normalsize



\section{Kinematics}

Estimates of the galaxy central velocity dispersion $\sigma_0$ and radial velocity were derived with the {\tt IRAF} task {\tt fxcor}. This program uses the cross-correlation method to measure the redshift of the galaxy spectra. A comparison between our redshift determination and that taken from {\tt NED} is presented in Figure 5-a. Simultaneously, the measured width (FWHM) of the correlation peak is used to estimate the velocity dispersion (FWHM = 2 $\sqrt{ln 4}$ $\sigma$). The width of the peak will be related to the broadening of the galaxy absorption lines compared to the instrumental resolution of the template star, i.e., FWHM$^2_{fxcor}$ = FWHM$^2_{galaxy}$ + FWHM$^2_{template}$.

Within the {\tt fxcor} task the spectra of galaxy and template are continuum subtracted and cut to a user defined wavelength range (4350-5250 [\AA] and 5300-6200 [\AA], for the blue and red observations respectively). Then cross-correlation is performed in the Fourier space, and a ``square'' filter is applied to remove large scale variations as well as noise from the spectra. The instrumental dispersion of the OAGH observations is approximately 100 km s$^{-1}$ pixel$^{-1}$ for the blue region and 80 km s$^{-1}$ pixel$^{-1}$ for the red region\footnote[2]{{\tt Note that to measure Lick indices and compare with literature data and models we have to degrade the resolution of the spectra to the values in Table 4. However, to measure velocity dispersion and radial velocity we have used our original observations where {\it no smoothing} of the data was performed.}}.  
Multiple measurements of the same galaxy were averaged. We have adopted the  error of the mean as our final velocity dispersion error (equation 7).

We note that the H$\beta$ feature, present in one of the wavelength intervals, is known to be a source for severe template mismatch for galaxies with strong Balmer absorption (Kuntschner 1998, 2000). Indeed we find some galaxies in our sample with very strong H$\beta$ feature, i.e., H$\beta$ $>$ 2.5 \AA\ (NGC~3156 has H$\beta$ $\simeq$ 3.5 \AA). For the same reason, we also have opted to exclude the H$\gamma$ feature from our velocity dispersion measurements. The red wavelength interval on the other hand, encompasses the NaD band, which has an important contribution from interstellar absorption (Dressler 1984). To avoid these template mismatch and contamination problems without losing important information from other wavelengths, we have removed the H$\beta$ and NaD features from the template spectra used for the cross-correlation. This way, the weight of the H$\beta$ and NaD regions during the cross-correlation in the Fourier space is kept lower in comparison to any of the other intense correlation features (e.g., the Mg triplet).

Figure 5 presents the comparison between our kinematical measurements and the literature. In panel (a) we see a good agreement (inside the errors) of our radial velocities and values obtained from NED. To do this comparison we have selected in NED the most recent radial velocities derived from optical spectroscopy.  
In Figure 5-b we show a comparison between our estimated central velocity dispersion and the mean scaled $\sigma_0$ in the compilation by Prugniel \& Simien (1996, PS96). In order to compare our measurements with literature values, we have corrected $\sigma_0$ to a standard aperture as described by J\o rgensen, Franx \& Kaj\ae rgaard (1995), which scales $\sigma$ to a normalized diameter equivalent to 3.4 arcsec projected on to a galaxy in the Coma cluster. This correction is generally $<$ 10 km/s for our sample. The results agree with those in the literature on a 70\% level, using a Kolmogorov-Smirnov test. For galaxies showing $\sigma_0 \la$ 100 km/s, systematic errors start to dominate as our spectral resolution is lower than the measurement, and therefore these low velocity dispersions should be considered only as rough estimates. Note that there is a small offset in Figure 5-b, our $\sigma_0$ is on average 16 km/s smaller than in PS96. 
For NGC~3379, we obtain $\sigma_0$ = 203 $\pm$ 7 km/s; Davies \etal (1987) find 201 $\pm$ 20 km/s; Franx \etal (1989): 220 $\pm$ 3 km/s; Tonry \& Davis (1981): 214 $\pm$ 15 km/s; Bender \etal (1994): 240 $\pm$ 5 km/s; and the average value adopted by Prugniel \& Simien (1996) for the compilation of 17 results on NGC~3379 is 221 km/s. 
We estimate the average uncertainty in the PS96 compilation to be at least of the order of 20 km/s from the multiple measurements, whereas our mean error bar is 11 km/s (or in logarithmic scale, our mean error is 0.024 in $\sigma$). 
Thus, we do not regard as important the offset between our data and the PS96 compilation.  

Table 7 lists the adopted radial velocities and aperture re-scaled central velocity dispersions for our galaxy sample.  

\begin{figure}
\vspace{10.2 cm}
\includegraphics{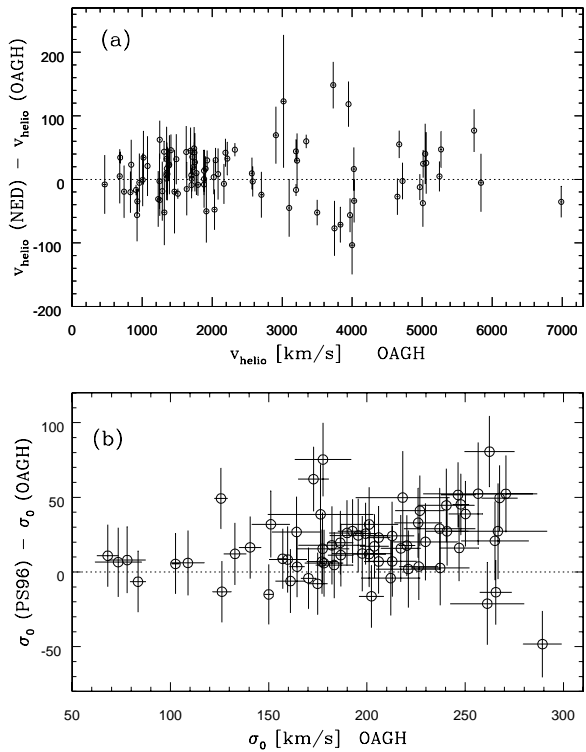}
\caption{(a) Comparison between our measured radial velocities and published values. (b) Comparison of our measured central velocity dispersions with the $\sigma_0$ from the compilation of Prugniel \& Simien (1996, PS96). The average error bar adopted for the PS96 compilation is 20 km/s. }
\end{figure}

\begin{table*}
\centering
\begin{minipage}{127mm}
\caption{Measured radial velocities (v$_{helio}$) and central velocity dispersions ($\sigma_0$) scaled to a standard circular aperture (see text, Section 5). }
\begin{tabular}{@{}lrr|clrr@{}}
\hline
\multicolumn{1}{l}{Galaxy}  &
\multicolumn{1}{r}{v$_{helio}$ [km/s]} &  
\multicolumn{1}{r}{$\sigma_0$ [km/s]}  &
\multicolumn{1}{c}\vline & 
\multicolumn{1}{c}{Galaxy}  &
\multicolumn{1}{r}{v$_{helio}$ [km/s]} &  
\multicolumn{1}{r}{$\sigma_0$ [km/s]} \\
\hline
ESO 462-G015 & 5840 $\pm$  27 & 239 $\pm$  24 & \vline & NGC 3599   &  832 $\pm$  ~9 &  68 $\pm$  ~5 \\
MCG -01-03-018&5743 $\pm$  10 & 191 $\pm$  14 & \vline & NGC 3607   & 960 $\pm$  20 & 221 $\pm$  16  \\ 
NGC 0016   & 3100 $\pm$  17 & 151 $\pm$  ~9   & \vline & NGC 3608   & 1229 $\pm$  12 & 182 $\pm$  16  \\
NGC 0221   & -123 $\pm$  10 &  83 $\pm$  ~4   & \vline & NGC 3610   & 1704 $\pm$  11 & 174 $\pm$  ~5  \\ 
NGC 0315   & 4968 $\pm$  11 & 246 $\pm$  ~8   & \vline & NGC 3613   & 2034 $\pm$  15 & 205 $\pm$  ~7 \\ 
NGC 0474   & 2325 $\pm$  ~7 & 159 $\pm$  ~9   & \vline & NGC 3636   & 1741 $\pm$  13 & 127 $\pm$  ~3  \\    
NGC 0584   & 1883 $\pm$  11 & 199 $\pm$  ~5   & \vline & NGC 3640   & 1251 $\pm$  12 & 178 $\pm$  ~9 \\
NGC 0720   &  1725 $\pm$  ~9 & 261 $\pm$  18  & \vline & NGC 3665   & 2049 $\pm$  10 & 202 $\pm$  ~6 \\   
NGC 0750   & 5248 $\pm$  16 & 189 $\pm$  ~9   & \vline & NGC 3923   & 1796 $\pm$  13 & 289 $\pm$  ~9 \\  
NGC 0751   & 5270 $\pm$  15 & 186 $\pm$  ~6   & \vline & NGC 3941   &  926 $\pm$  ~9 & 125 $\pm$  ~3  \\  
NGC 0777   & 5015 $\pm$  ~8 & 267 $\pm$  ~9   & \vline & NGC 4125   & 1347 $\pm$  35 & 201 $\pm$  14  \\
NGC 0821   & 1717 $\pm$  11 & 212 $\pm$  15   & \vline & NGC 4261   & 2191 $\pm$  11 & 256 $\pm$  27 \\    
NGC 0890   & 4027 $\pm$  26 & 205 $\pm$  ~5   & \vline & NGC 4365   & 1248 $\pm$  14 & 236 $\pm$  18 \\ 
NGC 1045   & 4646 $\pm$  ~9 & 230 $\pm$  16   & \vline & NGC 4374   & 1007 $\pm$  14 & 247 $\pm$  ~4 \\
NGC 1052   & 1510 $\pm$  ~6 & 176 $\pm$  26   & \vline & NGC 4494   & 1344 $\pm$  11 & 161 $\pm$  ~7 \\    
NGC 1132   & 6988 $\pm$  17 & 237 $\pm$  14   & \vline & NGC 4550   &  465 $\pm$  21 &  73 $\pm$  11  \\
NGC 1407   & 1774 $\pm$  19 & 265 $\pm$  17   & \vline & NGC 4754   & 1347 $\pm$  ~9 & 186 $\pm$  ~6  \\  
NGC 1453   & 3999 $\pm$  28 & 250 $\pm$  ~8   & \vline & NGC 5322   & 1754 $\pm$  16 & 217 $\pm$  10   \\ 
NGC 1600   & 4720 $\pm$  17 & 262 $\pm$  12   & \vline & NGC 5353   & 2169 $\pm$  16 & 266 $\pm$  25   \\   
NGC 1700   & 3971 $\pm$  13 & 220 $\pm$  ~4   & \vline & NGC 5354   & 2579 $\pm$  13 & 208 $\pm$  ~8   \\     
NGC 1726   & 4025 $\pm$  24 & 212 $\pm$  11  & \vline & NGC 5363   & 1077 $\pm$  ~9 & 175 $\pm$  33   \\
NGC 2128   & 3019 $\pm$  30 & 178 $\pm$  10  & \vline & NGC 5444   & 3948 $\pm$  10 & 212 $\pm$  12   \\  
NGC 2300   & 1905 $\pm$  ~7 & 265 $\pm$  ~8  & \vline & NGC 5557   & 3213 $\pm$  18 & 226 $\pm$  12   \\   
NGC 2418   & 5043 $\pm$  ~7 & 240 $\pm$  10  & \vline & NGC 5576   & 1487 $\pm$  ~6 & 183 $\pm$  ~9   \\ 
NGC 2513   & 4672 $\pm$  13 & 246 $\pm$  ~9  & \vline & NGC 5638   & 1637 $\pm$  11 & 157 $\pm$  ~4   \\   
NGC 2549   & 1019 $\pm$  14 & 140 $\pm$  ~5  & \vline & NGC 5812   & 1885 $\pm$  13 & 201 $\pm$  ~8    \\  
NGC 2768   & 1360 $\pm$  ~6 & 177 $\pm$  13  & \vline & NGC 5813   & 1929 $\pm$  ~7 & 226 $\pm$  ~9    \\   
NGC 2872   & 3196 $\pm$  11 & 240 $\pm$  14  & \vline & NGC 5831   & 1629 $\pm$  ~8 & 164 $\pm$  ~3    \\ 
NGC 2911   & 3199 $\pm$  ~7 & 172 $\pm$  ~7  & \vline & NGC 5845   & 1410 $\pm$  22 & 200 $\pm$  14   \\   
NGC 2974   & 1916 $\pm$  23 & 197 $\pm$  15  & \vline & NGC 5846   & 1708 $\pm$  17 & 219 $\pm$  14   \\  
NGC 3091   & 3729 $\pm$  17 & 285 $\pm$  22  & \vline & NGC 5846A  & 2217 $\pm$  16 & 183 $\pm$  ~8   \\  
NGC 3098   & 1387 $\pm$  13 & 108 $\pm$  ~8  & \vline & NGC 5854   & 1691 $\pm$  35 & 135 $\pm$  ~4   \\ 
NGC 3115   &  685 $\pm$  ~7 & 218 $\pm$  23  & \vline & NGC 5864   & 1885 $\pm$  ~4 & 130 $\pm$  ~5    \\ 
NGC 3139   & 1412 $\pm$  10 & 184 $\pm$  ~9  & \vline & NGC 5869   & 2085 $\pm$  ~4 & 162 $\pm$  18   \\
NGC 3156   & 1318 $\pm$  16 &  78 $\pm$  ~9  & \vline & NGC 5982   & 2911 $\pm$  22 & 229 $\pm$  16   \\     
NGC 3193   & 1377 $\pm$  18 & 220 $\pm$  20  & \vline & NGC 6172   & 5009 $\pm$  ~8 & 136 $\pm$  ~6   \\
NGC 3226   & 1287 $\pm$  11 & 164 $\pm$  12  & \vline & NGC 6411   & 3747 $\pm$  16 & 177 $\pm$  ~7    \\  
NGC 3245   & 1314 $\pm$  ~6 & 192 $\pm$  ~7  & \vline & NGC 7302   & 2703 $\pm$  16 & 188 $\pm$  ~7   \\    
NGC 3377   &  679 $\pm$  17 & 132 $\pm$  ~5  & \vline & NGC 7332   & 1250 $\pm$  16 & 150 $\pm$  ~4   \\    
NGC 3379   &  912 $\pm$  ~6 & 203 $\pm$  ~7  & \vline & NGC 7454   & 2022 $\pm$  ~5 & 126 $\pm$  ~4   \\ 
NGC 3384   &  740 $\pm$  15 & 170 $\pm$  ~3  & \vline & NGC 7585   & 3499 $\pm$  14 & 195 $\pm$  14   \\    
NGC 3412   &  844 $\pm$  10 & 102 $\pm$  ~4  & \vline & NGC 7619   & 3833 $\pm$  27 & 270 $\pm$  16   \\   
NGC 3414   & 1460 $\pm$  23 & 177 $\pm$  14  & \vline & NGC 7626   & 3345 $\pm$  ~9 & 226 $\pm$  12   \\
\hline
\end{tabular}
\end{minipage}
\end{table*}

\subsection{Velocity dispersion correction to the indices}

The observed spectrum of a galaxy is the convolution of the
integrated spectrum of its stellar population(s) with the
distribution of line-of-sight velocities of the stars and
instrumental broadening. These effects broaden the spectral features,
in general reducing the observed line-strength compared to intrinsic
values. In fact, it is well known that there is a velocity dispersion dependence on the indices and it needs to be corrected (e.g., Gonz\'alez 1993; J\o rgensen \etal 1995).
 In order to compare the raw index measurements in galaxies
with the model predictions we calibrate the indices to zero velocity
dispersion. The procedure adopted here is to broaden the spectra of
template stars with gaussians of $\sigma_0$ = 20-400 km/s in bins of
20 km/s. The indices are then measured for each $\sigma$ bin and a
correction factor is determined. The correction factor has the form
C($\sigma$) = index($\sigma$=0)/index($\sigma$) for indices measured
in \AA ngstroms, and C($\sigma$) = index($\sigma$=0)--index($\sigma$) for
indices given in magnitudes. It is important to stress that derived
correction factors are only useful if the stars used for the
simulations resemble the galaxy spectra. For this reason, we have
opted to build a composite stellar spectrum template, and compared the
resulting super-template with the spectra of NGC~3379,
observed in every run.  The super-template was created by averaging
different stellar-type  spectra selected from Table 3, carefully
allocating more weight  to the K giant stars. Figure 6
shows the dependence of  the correction factor on $\sigma$ for 25 Lick
indices measured in  stars; the dotted bars denote the uncertainty
range in the correction  due to errors in the index measurements. A
polynomial fitting to the  curves in Figure 6 allowed for the velocity
dispersion corrections,  which were then added or multiplied (if in
units of mag or \AA\  respectively) to the raw index measurements of
the resolution  corrected galaxy spectra. In this procedure, the
errors in $\sigma_0$ are then propagated to the index errors.

\begin{figure*}
\vspace{11.2 cm}
\includegraphics{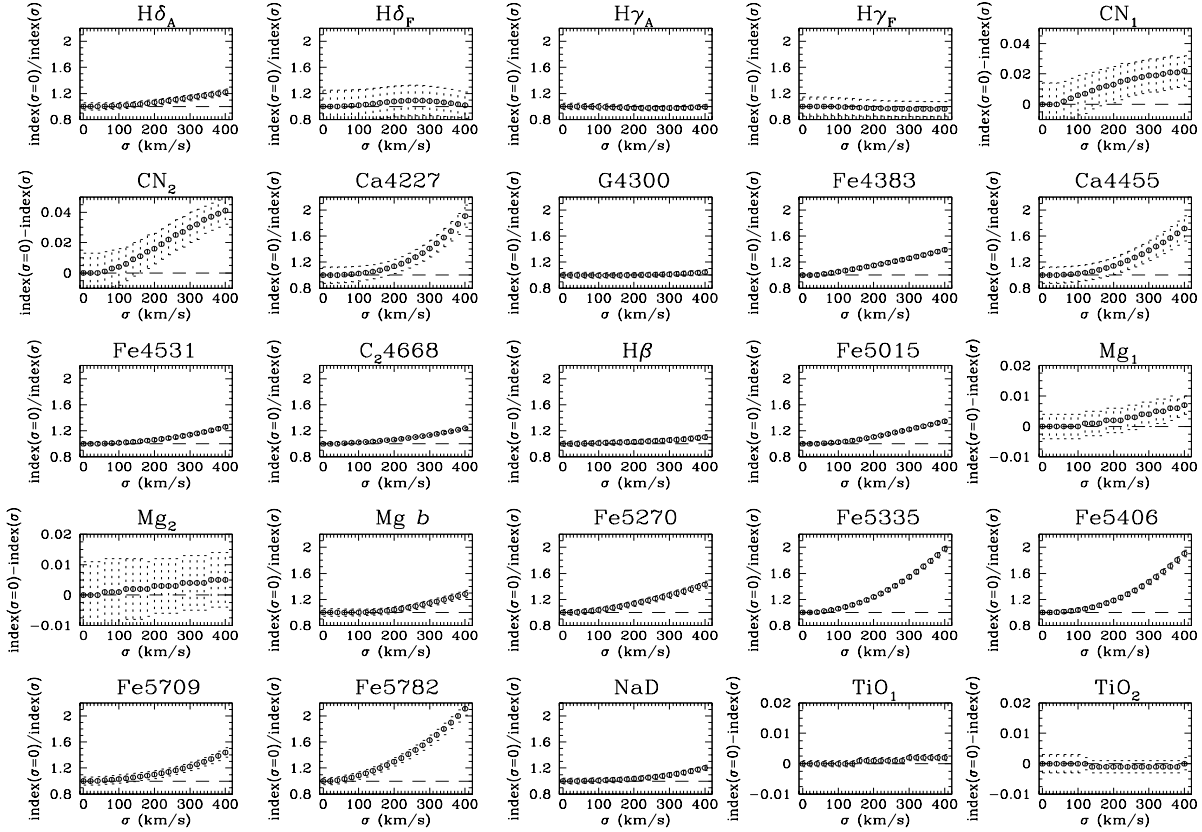}
\caption{Velocity dispersion corrections. The $\sigma$ broadening and 
index measurements were performed on a composite stellar spectrum 
 template that very well matches the spectra of the bona fide 
elliptical galaxy NGC~ 3379. The dotted bars represent the 
uncertainties derived from the index measurement errors. The dashed 
line is used as a visual guide-line for no correction applied.}
\end{figure*}

\begin{figure*}
\vspace{11.2 cm}
\includegraphics{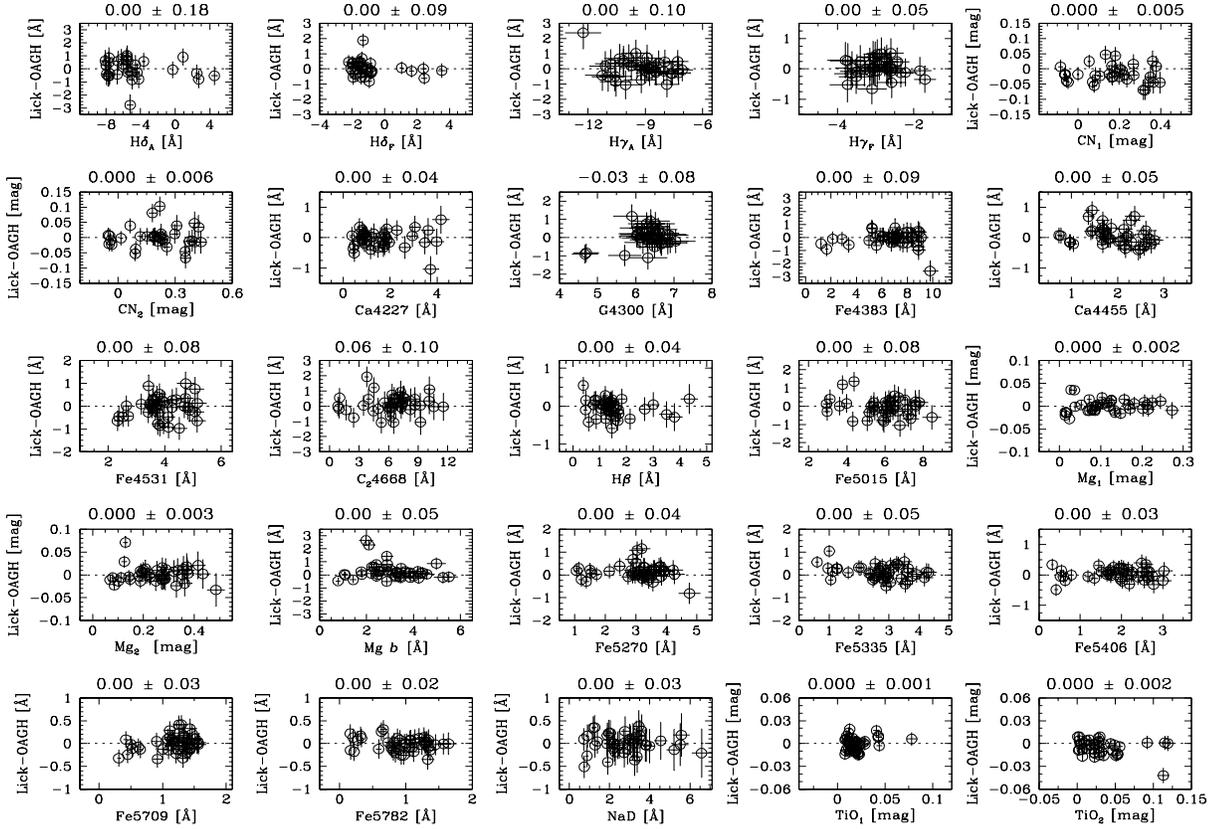}
\caption{Comparison between our line-strength measurements and Worthey \etal (1994) for 45 stars in common. Our data were corrected for the Lick/IDS offset prior to comparison. The remaining offset and associated 1$\sigma$ error are shown above each mini-plot. The offsets are summarised in Table 8.}
\end{figure*}



\section{Final central absorption line-strength}

To remove any remaining systematic offset between our measurement and 
the Lick/IDS, we have compared the measurements of 25 line-strength 
indices of our 45 {\it stars} observed in common with Lick. 
Generally there is good agreement between the two data sets, and only 
small offsets are found (see Figure 7, for an 
offset-corrected comparison). The mean offsets and associated errors 
for each index are summarized in Table 8. These offsets were 
then applied to our data. 

The final corrected central r$_e$/8 index measurements and associated 
errors for our galaxy sample are presented in the Appendix B in Table B.1.

\begin{table}
\centering
\begin{minipage}{50mm}
\caption{Lick/IDS offsets.}
\vskip 0.2cm
\begin{tabular}{@{}lc@{}}
\hline
\multicolumn{1}{l}{Index} &
\multicolumn{1}{c}{Offset (Lick/IDS $-$ OAGH)} \\
\hline
H$\delta_{A}$ & $-$0.07 $\pm$  0.18  \AA     \\ 
H$\delta_{F}$ & $-$0.09 $\pm$  0.09 \AA    \\
H$\gamma_{A}$ & +0.16 $\pm$  0.10 \AA    \\
H$\gamma_{F}$ & +0.18 $\pm$  0.05 \AA     \\
CN$_1$        &    0.000 $\pm$  0.005 mag   \\
CN$_2$        & $-$0.011 $\pm$  0.006 mag   \\
Ca4227        & $-$0.04 $\pm$  0.04 \AA    \\
G4300         & $-$0.03 $\pm$  0.08  \AA \\
Fe4383        & +0.19 $\pm$  0.09 \AA    \\
Ca4455        & +0.29 $\pm$  0.05 \AA     \\
Fe4531        & +0.27 $\pm$  0.08 \AA     \\
C$_{2}$4668   & +0.06 $\pm$  0.10 \AA     \\
H$\beta$      & $-$0.06 $\pm$  0.04 \AA    \\
Fe5015        & $-$0.17 $\pm$  0.08 \AA    \\
Mg$_1$        & +0.011 $\pm$  0.002 mag     \\
Mg$_2$        & +0.023 $\pm$  0.003 mag     \\
Mg{\it b}     & $-$0.09 $\pm$  0.05 \AA    \\
Fe5270        & $-$0.07 $\pm$  0.04 \AA     \\
Fe5335        & $-$0.12 $\pm$  0.05 \AA     \\
Fe5406        & $-$0.03 $\pm$  0.03 \AA     \\
Fe5709        & +0.05 $\pm$  0.03 \AA     \\
Fe5782        & +0.05 $\pm$  0.02 \AA    \\
NaD           & +0.09 $\pm$  0.03 \AA    \\
TiO$_1$       & +0.006 $\pm$  0.001 mag     \\
TiO$_2$       & +0.002 $\pm$  0.002 mag     \\
\hline
\end{tabular}
\end{minipage}
\end{table}

\subsection{Summary: the sample and Lick indices}

Trager \etal (1998) have demonstrated that index errors can masquerade as real trends in the determination of ages, metallicities and their correlations with velocity dispersion. 
Here we give special treatment to the H$\beta$-index because this index will be used in subsequent papers to infer ages and metallicities, using for example, H$\beta$-[MgFe] index diagrams.
It happens so that the information on H$\beta$ errors is commonly used to separate and select the best galaxy data in the literature. The sample used by Trager \etal (1998) had a typical error of $\delta$H$\beta$ = 0.191 \AA.
A reasonable guide is that H$\beta$ must be accurate to $\sim$ 0.1 \AA\ in order to determine reliable ages and metallicities. 

Figure 8 shows a histogram of H$\beta$-index errors. These errors take into account the errors in the emission correction, i.e., we have propagated the errors of the emission correction into the new H$\beta_{correct}$ error following eq. 9. The median error of the sample is 0.125 \AA. Previous to the emission correction, the median error of our data was 0.086~\AA.

The errors in the determination of ages and metallicities from index-index diagrams will be addressed again in a subsequent paper.

\begin{figure}
\vspace{8. cm}
\includegraphics{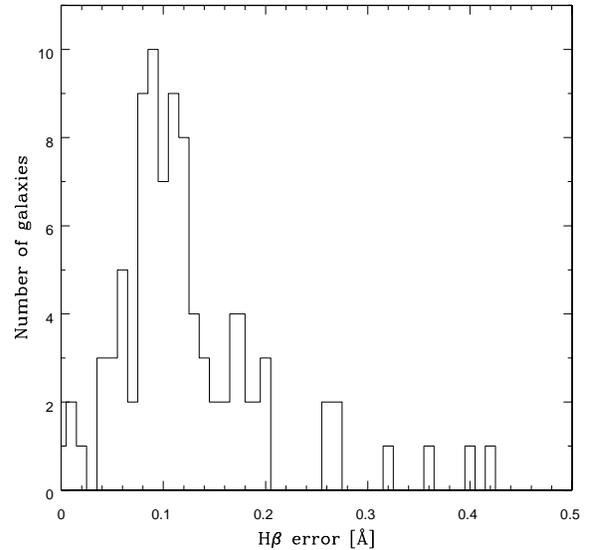}
\caption{ Histogram of H$\beta$-index errors (error of the mean per galaxy). These uncertainties take into account the errors in the emission correction. The median error of the sample is  0.125 \AA. }
\end{figure}

Our survey has the advantage of covering a large wavelength range,
from $\sim$ 3850 \AA\ to 6700 \AA, giving H$\alpha$ 
information which is important for emission correction. 
Of the 86 galaxies with S/N $\geq$ 15 (for r$_e$/8 aperture, and per resolution element), 52 had 
H$\beta$-index corrected for emission, the corrections varied from 
 as small as $\sim$ 0.1 \AA\ to as large as
$\sim$ 3.3 \AA\ for NGC~1052; 41 galaxies present
\Oiii$\lambda$5007 emission, of which 16 also show obvious 
H$\alpha$ emission. Most of the galaxies in the sample do not show 
obvious signs of disturbances nor tidal features in their
morphologies, although 11 galaxies belong to the Arp catalogue of 
peculiar galaxies (Arp 1966), of which only three (NGC~750, NGC~751 and 
NGC~3226) seem to be strongly interacting.

We have applied a better method for emission correction of 
the Balmer-line indices than the use of the uncertain 
 0.6$\times$\Oiii estimation. The new correction uses the 
intensity and equivalent width of the H$\alpha$ index (defined in Table 5).
We note that nebular emission could still be affecting the ages and 
metallicities derived from most of the data in the literature.



\section{Index--$\sigma$ relations}

In this Section, we will compare log $\sigma_0$ with indices measured in magnitudes. Indices originally measured in \AA ngstroms are converted to magnitudes as explained in eq. 4, and now denoted with a prime, i.e., H$\beta$ in magnitudes will be called H$\beta^{\prime}$.

The dynamical properties of galaxy cores are closely connected with their stellar populations, implied by the relatively small scatter in the colour-$\sigma_0$, Mg-$\sigma_0$ relations found in previous investigations (e.g.~Terlevich \etal 1981, Burstein \etal 1988, Bower, Lucey \& Ellis 1992). The prominence of the Mg-$\sigma_0$ relation suggests that other metal line-strength indices should also exhibit a correlation with the central velocity dispersion. We present in Figures 9 and 10 the index-$\sigma_0$ relations for $<$Fe$>^{\prime}$, CN$_1$, two Balmer-line indices (H$\beta^{\prime}$, H$\gamma_A^{\prime}$), Mg{\it b}$^{\prime}$ and Mg$_2$.

We have performed ordinary least square fits with Y = Index$^{\prime}$, as the dependent variable (hereafter the (Y$\mid$X) fit; solid line in Figures 9 and 10) and also a fit in which Index$^{\prime}$ is the independent variable (dotted line in Figures 9 and 10). Isobe \etal (1990) recommend the (Y$\mid$X) fit for scientific problems where one variable is clearly an effect and the other the cause. The combination of the regressions of Y on X and X on Y allows us to compute the linear correlation coefficient R = $\sqrt{{\it b} {\it b}^{\prime} }$, where Y = {\it a} + {\it b}X and X = {\it a}$^{\prime}$ + {\it b}$^{\prime}$Y. Here both regressions are shown to illustrate how different fitting methods lead to different results. For the Index$^{\prime}$--$\sigma_0$ relation analysis, we favour the (Y$\mid$X) fit method for two reasons: (1) consistency with previous authors, as frequently they fit their data using the (Y$\mid$X) method; (2) at least the Mg content of a galaxy seems to be driven by the central potential of the galaxy (see conclusions of Colless \etal 1999) which is approximated by its central velocity dispersion $\sigma_0$. The code used to derive the (Y$\mid$X) fits is an {\tt IRAF} implementation of the Fortran code by Bevington (1969). Table 9 presents the (Y$\mid$X) fit values and uncertainties for all the sample (Es+S0s).

\begin{table*}
\begin{minipage}{100cm}
\caption{Least square fit results. }
\vskip 0.05cm
\begin{tabular}{@{}rcrcccc@{}}
\hline
\multicolumn{1}{r}{Index} &
\multicolumn{1}{c}{} &
\multicolumn{1}{c}{} &
\multicolumn{1}{c}{$<$$\Delta$Index$^{\prime}$$>$} &
\multicolumn{1}{c}{$<$$\delta$Index$^{\prime}$$>$} &
\multicolumn{1}{c}{$<$$\delta$i$>$} &
\multicolumn{1}{c}{R$^{\dagger}$} \\
\multicolumn{1}{r}{} &
\multicolumn{1}{l}{} &
\multicolumn{1}{c}{} &
\multicolumn{1}{c}{[mag]} &
\multicolumn{1}{c}{[mag]}  &
\multicolumn{1}{c}{[mag]} &
\multicolumn{1}{c}{}\\
\hline

Mg$_2$       &= & $-$0.198($\pm$0.048)+0.218($\pm$0.021)$\cdot$ log\ $\sigma_0$ & 0.024 & 0.003 & 0.023&0.75\\
Mg{\it b}$^{\prime}$   &= & $-$0.277($\pm$0.012)+0.190($\pm$0.005)$\cdot$ log\ $\sigma_0$ & 0.016 & 0.004 & 0.016&0.74\\
$<$Fe$>$$^{\prime}$&= & 0.003($\pm$0.014)+0.044($\pm$0.006)$\cdot$ log\ $\sigma_0$ & 0.007 & 0.005 & 0.004&0.61\\
CN$_1$       &= & $-$0.457($\pm$0.069)+0.244($\pm$0.030)$\cdot$ log\ $\sigma_0$ & 0.036 & 0.006 & 0.035&0.61\\
H$\beta$$^{\prime}$    &= & 0.267($\pm$0.028)$-$0.086($\pm$0.012)$\cdot$ log\ $\sigma_0$  & 0.012 & 0.003 & 0.011&0.55\\
H$\gamma_A$$^{\prime}$ &= & 0.110($\pm$0.033)$-$0.108($\pm$0.015)$\cdot$ log\ $\sigma_0$  & 0.016 & 0.004 & 0.015&0.63\\

\hline
\end{tabular}
\vskip 0.1cm

{$\dagger$ R is the linear correlation coefficient calculated as R = $\sqrt{{\it b} {\it b}^{\prime}}$, where \\
log $\sigma_0$ = {\it a} + {\it b} Index$^{\prime}$, and Index$^{\prime}$ = {\it a}$^{\prime}$ + {\it b}$^{\prime}$ log $\sigma_0$. The correlation R varies from 0 to 1.}

\noi{$<$$\Delta$Index$^{\prime}$$>$: Average deviation from fit (excluding NGC~3156).}

{$<$$\delta$Index$^{\prime}$$>$: Average individual error in the index.}

{$<$$\delta$i$>$: Estimated intrinsic scatter (excluding NGC~3156).}

\end{minipage}
\end{table*}

In Figures 9 and 10 we compare our results with the Fornax cluster sample from Kuntschner (2000; 22 galaxies), and the `mixed' galaxy sample from Kuntschner \etal (2001; 72 galaxies), which includes Virgo and Coma clusters galaxies, a few S0s and galaxies in less dense environments. Consistent with the results of Kuntschner (2000) and Kuntschner \etal (2001, 2002), we find a weak correlation between the $<$Fe$>$ index and $\sigma_0$. The slope of the $<$Fe$>$$^{\prime}$ -- log $\sigma_0$ relation is the smallest in Table 9. The Mg$_2$--$\sigma_0$ and $<$Fe$>$$^{\prime}$--$\sigma_0$ relations presented here for group, field and isolated galaxies are not significantly different from those of cluster E/S0s (see fits plotted in Figures 9 and 10). Nonetheless, the slopes of the other relations (except CN$_1$ -- log $\sigma_0$, which has no comparison) are significantly different with respect to previous authors, even though all relations are still following the trend of increasing metallicity with increasing velocity dispersion. 

It is necessary to point out that the shift in the dashed lines between the 
fit of $<$Fe$>$ and Mg{\it b}$^{\prime}$ indices for K2000 and K2001 in Figures 9(b) and 9(c) 
is probably related to 
the fact that, while K2000 are ``central'' values for the equidistant Fornax 
cluster galaxies, K2001 values are for a fixed linear aperture as in this 
paper.
We have no explanation for the shift in $<$Fe$>$ indices between K2001 and 
this work, other than that K2001 includes a more ``mixed'' sample of 
galaxies.

The fit values for the Fornax cluster in Kuntschner (2000) and for the field, 
group and cluster galaxies in  Kuntschner \etal (2001) are shown in Table 10. 

\begin{table}
\begin{minipage}{100cm}
\caption{Kuntschner \etal (2000, 2001) least square fit results.}
\vskip 0.05cm
\begin{tabular}{@{}rcr@{}}
\hline
K2000:  & & \\
Mg$_2$       &= & $-$(0.127 $\pm$ 0.054) + (0.191 $\pm$ 0.023) log $\sigma_0$ \\
Mg{\it b}$^{\prime}$&= & $-$(0.056 $\pm$ 0.044) + (0.102 $\pm$ 0.020) log $\sigma_0$ \\
$<$Fe$>$$^{\prime}$ &= & $+$(0.015 $\pm$ 0.053) + (0.038 $\pm$ 0.023) log $\sigma_0$ \\
H$\beta$$^{\prime}$ &= & $+$(0.106 $\pm$ 0.015) $-$ (0.020 $\pm$ 0.007) log $\sigma_0$\\
H$\gamma_A$$^{\prime}$ &= & $-$(0.038 $\pm$ 0.044) $-$ (0.045 $\pm$ 0.019) log $\sigma_0$ \\
\hline
K2001:  & & \\
Mg{\it b}$^{\prime}$   &= & $-$(0.163 $\pm$ 0.031) + (0.142 $\pm$ 0.013) log $\sigma_0$ \\
$<$Fe$>$$^{\prime}$&= & $-$(0.034 $\pm$ 0.015) + (0.021 $\pm$ 0.006) log $\sigma_0$ \\
\hline
\end{tabular}
\vskip 0.1cm
\end{minipage}
\end{table}

Figures 9-d and 10-a show the index--$\sigma_0$ relations for two Balmer-line indices (H$\beta$$^{\prime}$ and H$\gamma_A$$^{\prime}$). Both indices show negative correlations with the central velocity dispersion.
In  Figure 10-b, the CN$_1$ index correlates with $\sigma_0$ almost as strongly as Mg$_2$, however with larger scatter. In a subsequent paper we will discuss the sensitivity of CN indices to $\alpha$-elements like Mg. 

Note that on average most S0 galaxies have slightly lower $\sigma_0$ values than the bulk of Es. Remarkably, bulges of S0s follow the same general relation with $\sigma_0$ as the elliptical galaxies. Yet, the peculiar S0 galaxy NGC~3156 shows a Mg absorption which is too low by 0.080 and 0.040 mag for Mg$_2$ and $<$Fe$>$$^{\prime}$ respectively. As we will in a subsequent paper, this galaxy has a very young (luminosity weighted) central stellar population for an early-type galaxy of its brightness and velocity dispersion. This may be an extreme case, 
but it demonstrates nicely how the Mg, Fe, H$\beta$, H$\gamma$ -- $\sigma_0$ relations can be influenced by young stellar populations in the centre of the galaxy. We also note that the bulk of galaxies in our sample with higher H$\beta$ and H$\gamma$ index values (i.e., indicative of younger stellar populations according to single-stellar population models [forthcoming paper]) have relatively small velocity dispersions. We note also that of the four galaxies with log $\sigma$ $<$ 2.0 in our sample, two are classified as field galaxies (NGC~3599, NGC~4550), and the other two are NGC~221, a compact dwarf elliptical, and NGC~3156, an outlier galaxy in our sample (c.f., Figures 9 and 10).

On average our Mg-absorption strength is lower than in the Fornax galaxies from Kuntschner (2000). This could either be a metallicity or/and age related effect. As we have seen, there is a general trend that the presence of young stellar populations moves galaxies to lower Mg values. Some examples of anomalously low Mg-absorption values were previously identified by J\o rgensen (1997). In agreement with this trend, we can also argue that our H$\beta$$^{\prime}$ values are slightly higher (i.e., younger) than the cluster sample.
This may be one indication that on average our OAGH sample is relatively younger in comparison to the mean age of galaxies in Fornax.

\begin{table*}
\begin{minipage}{180cm}
\caption{Correlation matrix of the residuals ($\Delta$) of the log $\sigma_0$-Index relations.}
\vskip 0.05cm
\begin{tabular}{@{}rcccccc@{}}
\hline
\multicolumn{1}{r}{} &
\multicolumn{1}{c}{$\Delta$Mg$_2$} &
\multicolumn{1}{c}{$\Delta$Mg{\it b}$^{\prime}$} &
\multicolumn{1}{c}{$\Delta$$<$Fe$>$$^{\prime}$} &
\multicolumn{1}{c}{$\Delta$CN$_1$} &
\multicolumn{1}{c}{$\Delta$H$\beta$$^{\prime}$} &
\multicolumn{1}{c}{$\Delta$H$\gamma_A$$^{\prime}$} \\
\multicolumn{1}{r}{} &
\multicolumn{1}{c}{[mag]} &
\multicolumn{1}{c}{[mag]} &
\multicolumn{1}{c}{[mag]} &
\multicolumn{1}{c}{[mag]} &
\multicolumn{1}{c}{[mag]} &
\multicolumn{1}{c}{[mag]} \\
\hline

M$_B$         [mag]& 0.17 & 0.40 & 0.11 & 0.17 & 0.02 & 0.22 \\
log $\sigma_0$[mag]& 0.08 & 0.27 & 0.19 & 0.03 & 0.08 & 0.36 \\
Mg$_2$        [mag]& 0.66 & 0.24 & 0.48 & 0.46 & 0.39 & 0.66 \\
Mg{\it b}$^{\prime}$    [mag]& 0.57 & 0.38 & 0.46 & 0.39 & 0.42 & 0.64 \\
$<$Fe$>$$^{\prime}$ [mag]& 0.45 & 0.17 & 0.84 & 0.32 & 0.43 & 0.78 \\
CN$_1$        [mag]& 0.59 & 0.26 & 0.40 & 0.71 & 0.31 & 0.52 \\
H$\beta$$^{\prime}$     [mag]& 0.47 & 0.21 & 0.47 & 0.30 & 0.82 & 0.78 \\
H$\gamma_A$$^{\prime}$  [mag]& 0.51 & 0.22 & 0.67 & 0.32 & 0.62 & 0.94 \\

\hline
\end{tabular}
\vskip 0.1cm
{\footnotesize{Note: Linear correlation coefficients (R) calculated as R = $\sqrt{{\it b} {\it b}^{\prime}}$, where \\
log $\sigma_0$ = {\it a} + {\it b} Index$^{\prime}$, and Index$^{\prime}$ = {\it a}$^{\prime}$ + {\it b}$^{\prime}$ log $\sigma_0$.\\ 
The correlation R varies from 0 to 1.}}
\end{minipage}
\end{table*}

The bottom panels in Figures 9 and 10 show the residuals from the Index$^{\prime}$-$\sigma_0$ fits (Y$\mid$X). The dashed lines indicate the 1-$\sigma$ intrinsic scatter; that is +0.016 and +0.004 mag for Mg{\it b}$^{\prime}$ and $<$Fe$>$$^{\prime}$ respectively.  We have estimated the intrinsic scatter by requiring 

\begin{equation}
\sum^N_{k=1}{\Delta Index^{\prime}{^2_k} \over {\delta Index^{\prime}{^2_k} + \delta i^2}} \equiv 1
\end{equation}

\noi{where $\Delta$Index$^{\prime}$$_k$ is the deviation from the fitting relation for each galaxy, $\delta$Index$^{\prime}$$_k$ are the individual errors in Index$^{\prime}$, and $\delta$i is the estimated intrinsic scatter. The values for the intrinsic scatter are shown in the last column of Table 9. Comparing with previous results, Colless \etal (1999) analysed 736 early-type galaxies from their EFAR cluster galaxy sample with an intrinsic scatter for Mg{\it b}$^{\prime}$ of 0.016 mag. Kuntschner \etal (2001), for a sample of 72 early-type galaxies from different environmental regions, derived an intrinsic scatter of 0.012 mag in Mg{\it b}$^{\prime}$, smaller than ours by 0.004 mag.

In Figure 9-a, the Mg$_2$-$\sigma_0$ relation, we also show the least square bisector fit by Guzm\'an \etal (1992) from a sample of 51 Coma cluster ellipticals. The Coma cluster is the densest cluster in the local universe (at average distance $\sim$ 7000 km/s).
The bisector fit by Guzm\'an \etal (1992) is

\begin{equation}
Mg_2 = - (0.316 \pm 0.003) + 0.260 \ log \sigma_0
\end{equation}

The {\it observed} scatter for their Mg$_2$-$\sigma$ relation is 0.020 mag.  The {\it intrinsic} scatter is  0.016 $\pm$ 0.002 mag.  Guzm\'an \etal's fit is very similar to the results of Colless \etal (1999) for the EFAR sample of cluster galaxies. Colless \etal (1999) maximum likelihood fit for Mg$_2$-$\sigma$ relation of 423 cluster early-type galaxies is Mg$_2$ = -(0.305 $\pm$ 0.064) + (0.257 $\pm$ 0.027) log $\sigma_0$. 

Note that the slope of our fit for the Mg$_2$-$\sigma_0$ relation is similar to the slopes of the EFAR and Coma cluster samples. However, comparing only the slopes and scatter of the relations is not telling us about important differences between the samples: we would need to analyse the distribution of the galaxies in the Mg-$\sigma$ relations to infer further conclusions about the different environment (cluster, field). Indeed, most of the galaxies in the EFAR sample are clustered around log $\sigma$ $\sim$ 2.35-2.40, and go beyond log $\sigma$ = 2.5, which is a slightly different range of $\sigma$ values from that covered by our group, field and isolated galaxies. We can do a better comparison with the Coma cluster sample of Guzm\'an \etal (1992) as they list the velocity dispersions and Mg$_2$ index strength for their sample.  The average log $\sigma_0$ in Guzm\'an \etal's sample is  2.32 (error of the mean: 0.02, for a sample of 51 galaxies), whereas in our sample the average value is log $\sigma_0$ = 2.302 (error of the mean: 0.014, for a sample of 86 galaxies). The average Mg$_2$ index strength in the Coma cluster sample is 0.282 (error of the mean: 0.004)  and in our sample is 0.307 (error of the mean: 0.004). Thus, the distribution of galaxies in the Mg$_2$-$\sigma_0$ relation for high or less dense regions is very similar.
We learn from these comparisons that, perhaps surprisingly, cluster, group and field galaxies have very similar Mg-$\sigma$ relations, and environment does not seem to be a key parameter here. It is relevant to note here that Bernardi \etal (1998) also found, for a large sample of 931 early-type galaxies, that objects assigned to cluster, group or field follow almost identical Mg$_2$-$\sigma_0$ relations.

We note that age and metallicity may conspire to keep the
relation tight (Trager et al. 2000b) and hence its
usefulness as a probe of different galactic environments may be
limited.

\

What causes the spread in the Index$^{\prime}$-$\sigma_0$ relations?

Two obvious potential sources of scatter are age and metallicity. 
In the case of the Mg-$\sigma_0$ relation, it has been investigated by many authors, but perhaps the best paper on this topic is by Bender \etal (1993). They conclude that if the spread in the Mg-$\sigma_0$ relation at a given $\sigma_0$ is due to age only then the rms spread in age is only 15\%  for bright dynamically hot galaxies. Alternatively, if the spread is only due to metallicity then they infer a rms spread of about 15\% in metallicity. A similar analysis by Colless \etal (1999) using up-to-date model predictions and constraints from the Fundamental Plane find slightly larger numbers. 

With the help of stellar population models one can translate the intrinsic spread in Mg{\it b}$^{\prime}$ for example, at a given $\sigma$, into age and metallicity spreads. To simplify this exercise, we assume initially that there are no other sources of scatter, and that age and metallicity are not correlated, or only mildly so (see Colless \etal 1999 for more detailed analysis). Using Colless \etal (1999) calibration of Mg{\it b}$^{\prime}$ as a function of log age and metallicity, we find for our data set at a given $\sigma$ and at fixed metallicity a spread of $\delta${\it t}/{\it t} = 67   per cent in age and at fixed age a spread of $\delta${\it Z}/{\it Z} = 43 per cent respectively. Kuntschner \etal (2001) found 49 and 32 per cent respectively.

In Table 11 we investigate the effects that age (in the form of the Balmer-line indices, H$\beta$, H$\gamma$, for example), metallicity (using Mg, Fe, as metallicity indicators), absolute magnitude (M$_B$), and $\sigma_0$ may have on  the scatter of the Index-$\sigma_0$ relations.  
Table 11 presents the results of the linear correlations coefficients (R) from the residuals ($\Delta$) {\it vs} M$_B$, log $\sigma_0$, Mg, Fe, H$\beta$, H$\gamma$ and CN indices. 

In general the residuals of the Index--$\sigma_0$ relations better correlate with the  strength of the Index itself, i.e., $\Delta$Mg$_2$ shows R = 0.66 for a correlation with Mg$_2$, $\Delta$H$\beta$$^{\prime}$ shows R = 0.82 for a correlation with H$\beta$$^{\prime}$, and so on. For the Mg$_2$-$\sigma_0$, the {\it scatter} of the relation grows with decreasing Mg$_2$ strength.  In the case of H$\beta$$^{\prime}$--$\sigma_0$, the scatter grows with increasing H$\beta$ strength. Note however that our sample, like most of the currently available samples, is not complete and is still lacking low-luminosity and low-$\sigma$ galaxies. Another work by Concannon, Rose \& Caldwell (2000) indicates that the spread in age at a given $\sigma$ increases towards the low velocity dispersion range. Hence incompleteness at low velocity dispersions is potentially a source for bias. Finally, the correlation of the residuals with H$\beta$ index is not specially significant to clearly associate the intrinsic scatters in our sample with possible age variations. The same weak correlations are observed for metal-line indices {\it vs} scatter, for our sample.

There may well be other effects responsible for the scatter, such as variations in the Mg-overabundance at a given $\sigma_0$. We leave further discussion to a forthcoming paper where we investigate the relations of $\sigma_0$ with age, metallicity and $\alpha$/Fe ratios and the scatter of the Mg--$\sigma_0$ relation with age and metallicity. 

Finally, we would like to comment on the work of Worthey and Collobert (2003), using a compilation of nearly 2000 Mg-$\sigma$ relations from the literature to assess the question of the importance of mergers in the formation of early-type galaxies. Their simulations suggest that the evolution of median Mg index strength is not a good discriminator between mergers and passive evolution and that better discriminator such as Mg-$\sigma$ scatter and asymmetry require samples of more than 1000 objects with accuracies similar to today's local measurements to really establish further constraints in the formation scenario of early-type galaxies.

\begin{figure*}
\vspace{21.8 cm}
\includegraphics{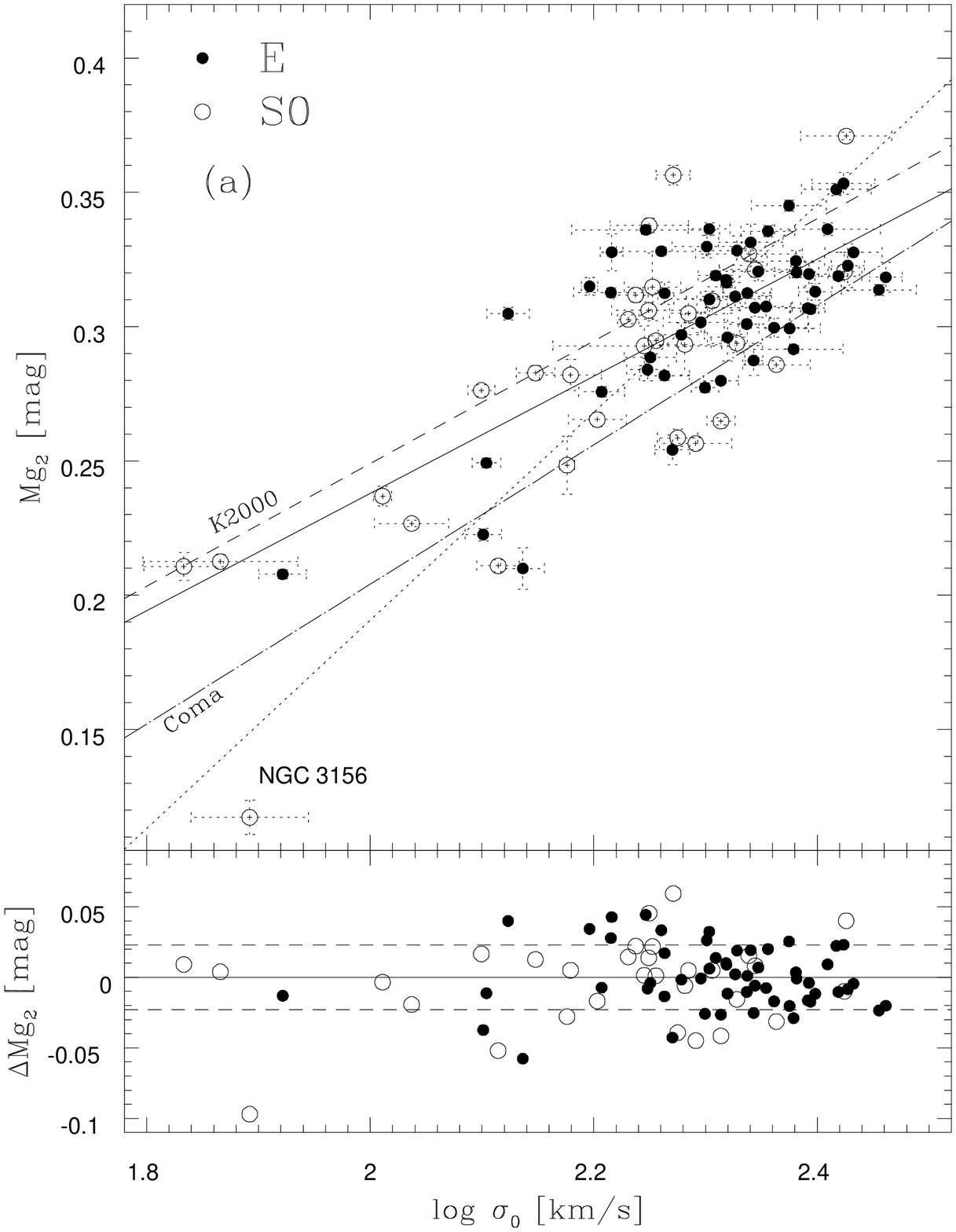}
\includegraphics{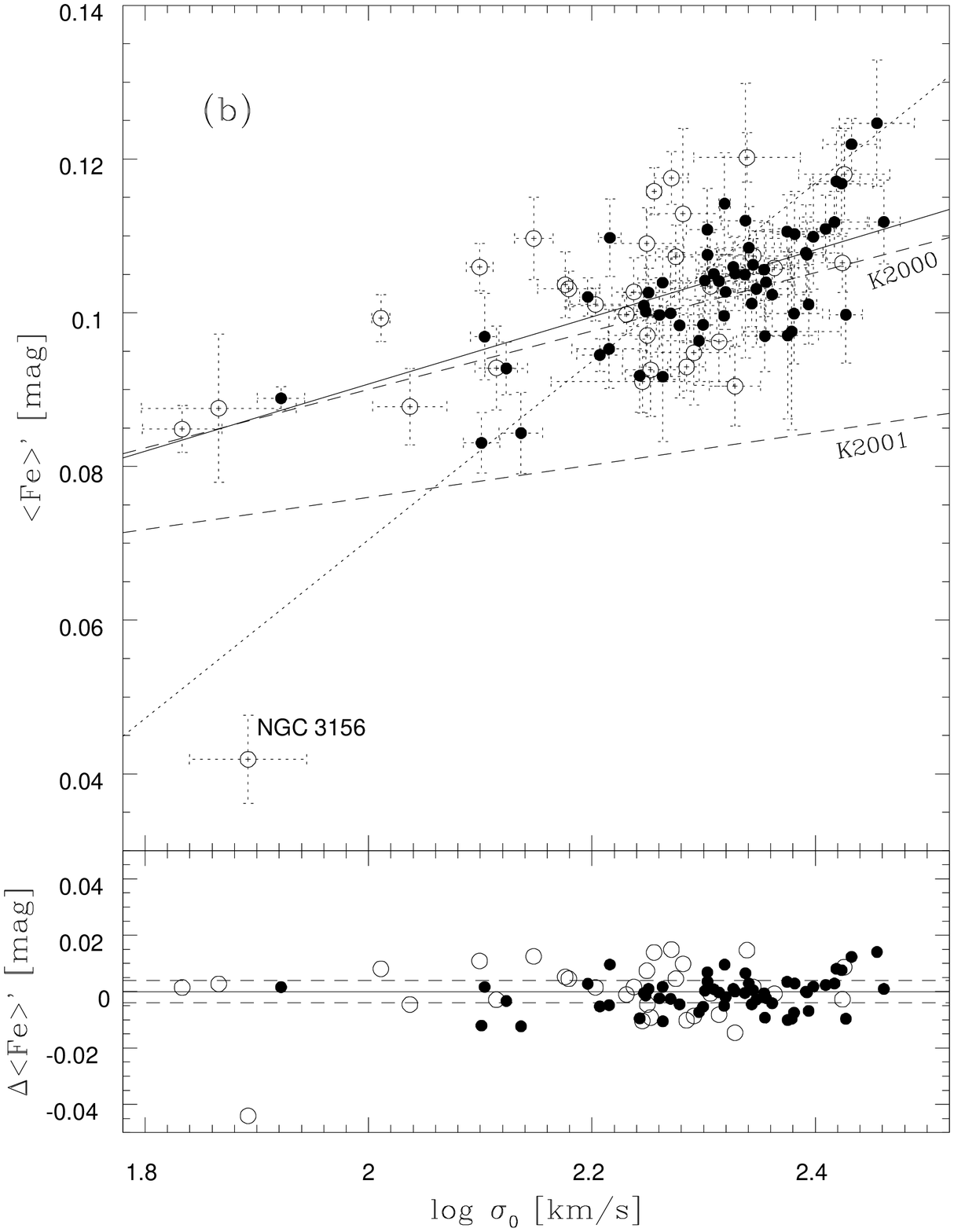}
\includegraphics{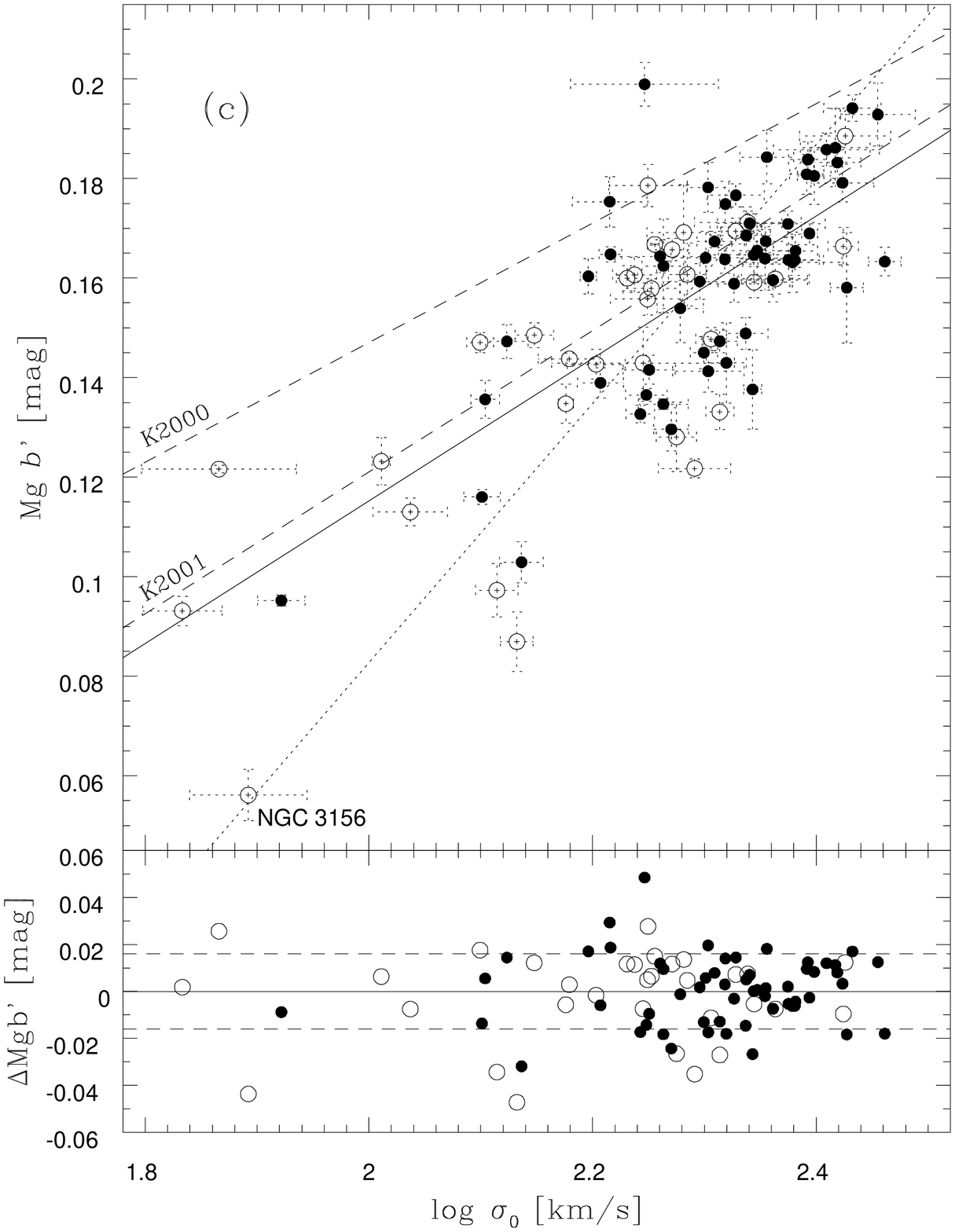}
\includegraphics{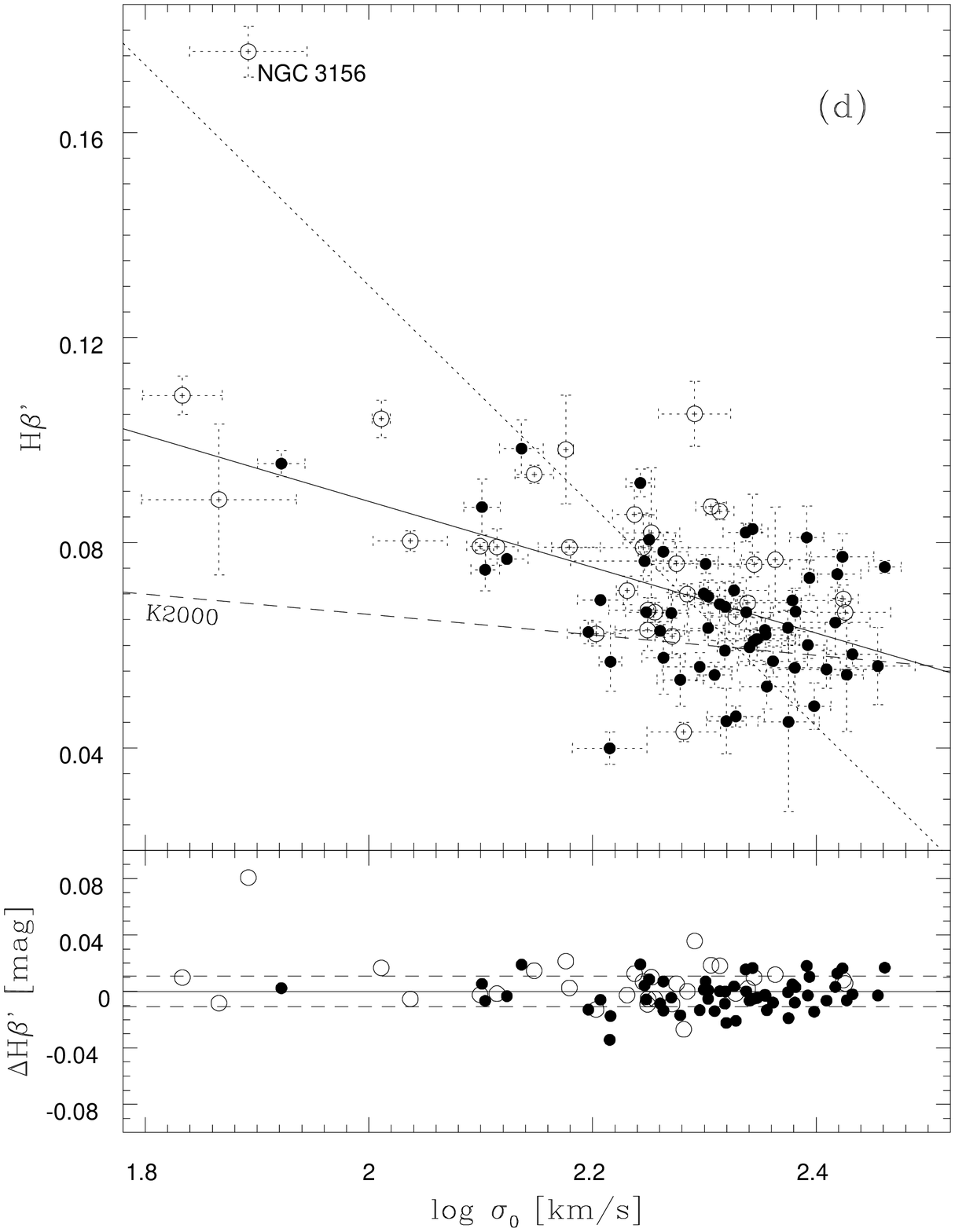}
\caption{(a) Mg$_2$ {\it vs} log $\sigma_0$; (b) $<$Fe$>$$^{\prime}$  {\it vs} log $\sigma_0$; (c) Mg{\it b}$^{\prime}$ {\it vs} log $\sigma_0$; (d) H$\beta$$^{\prime}$ {\it vs} log $\sigma_0$  relations. The indices are measured in magnitudes.
The dashed line indicates the fit for the cluster sample of Kuntschner (2000, K2000) and the cluster+field sample of Kuntschner \etal (2001, K2001).
The dot-dashed line in panel (a) represents the least square bisector fit by Guzm\'an \etal (1992) using a sample of 51 Coma cluster ellipticals.
Two fits are shown for our galaxy sample: the solid line is a normal least square fit with the index as dependent variable, and the dotted line is the ordinary least square fit with log $\sigma_0$ as dependent variable. 
The residuals of the relations, shown in the lower panels, are calculated with respect to the solid line fits. The dashed lines in the bottom panels indicate the 1-$\sigma$ intrinsic scatter in the residuals.}
\end{figure*}

\begin{figure*}
\vspace{11.2 cm}
\includegraphics{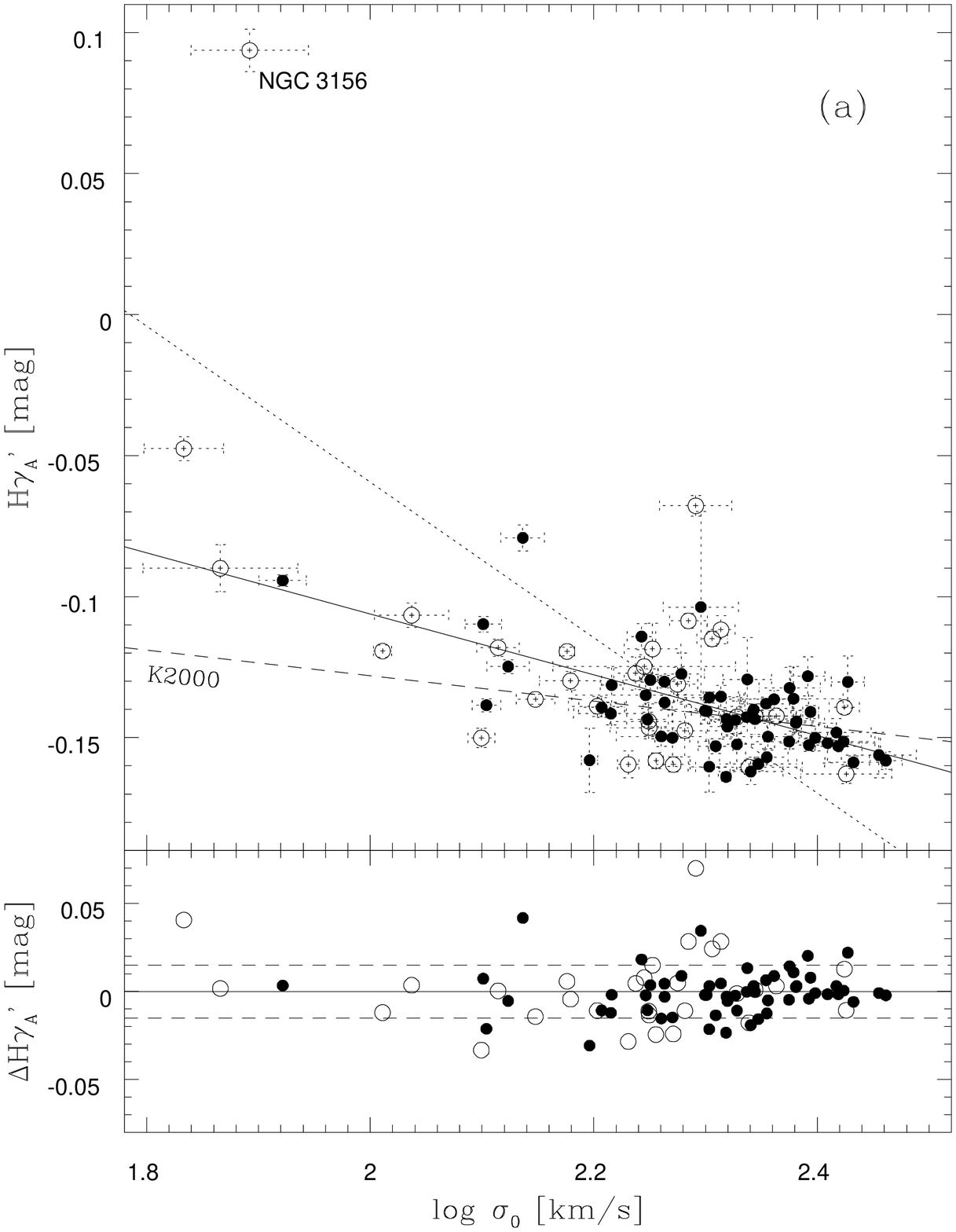}
\includegraphics{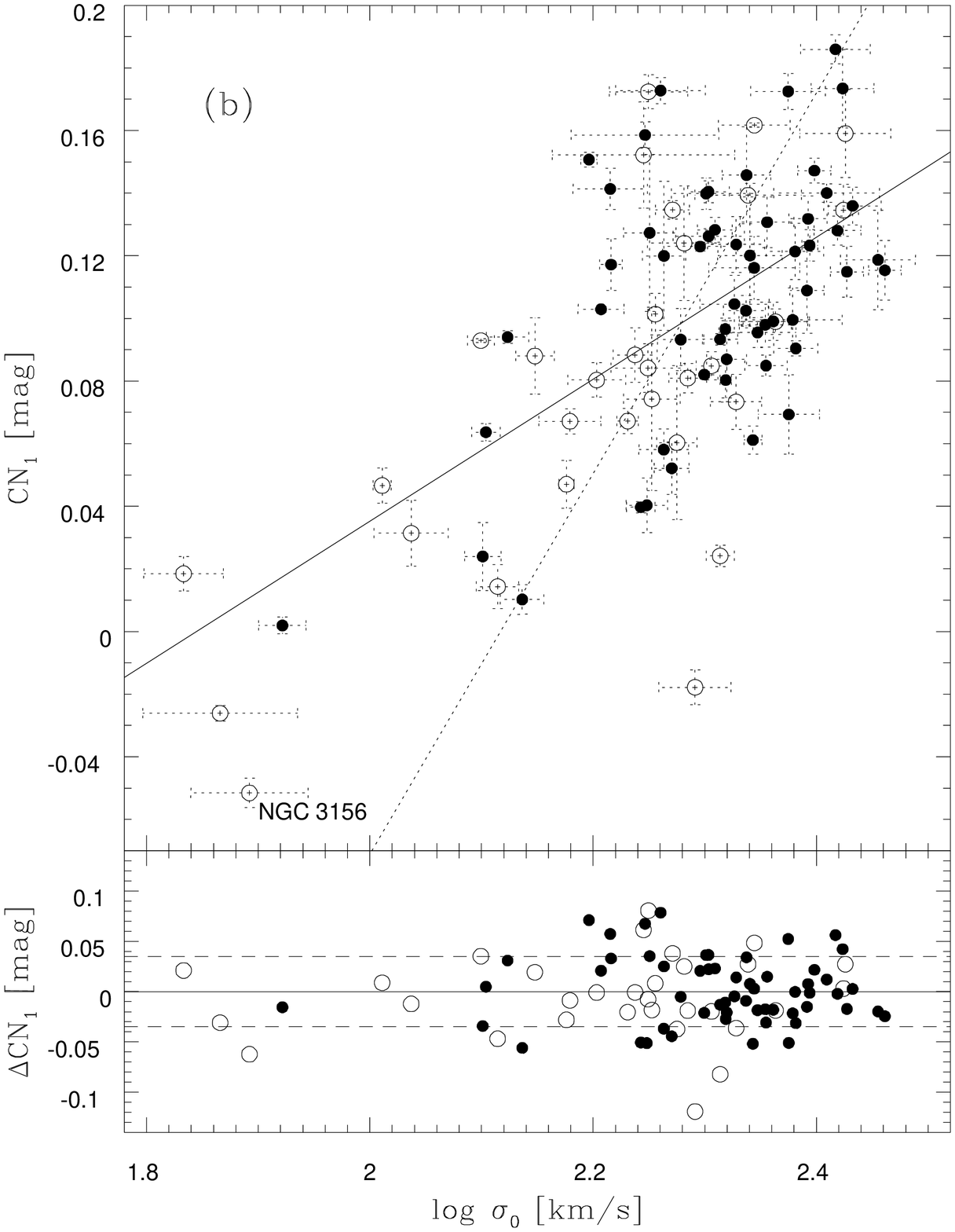}
\caption
{(a) H$\gamma_A$$^{\prime}$ {\it vs} log $\sigma_0$; (b) CN$_1$ {\it vs} log $\sigma_0$  relations. The indices are measured in magnitudes. The dashed line indicates the fit for the cluster sample of Kuntschner (2000, K2000). 
Two fits are shown for our galaxy sample: the solid line is a normal least square fit with the index as dependent variable, and the dotted line is the ordinary least square fit with log $\sigma_0$ as dependent variable.
The residuals of the relations, shown in the lower panels, are calculated with respect to the solid line fits. The dashed lines in the bottom panels indicate the 1-$\sigma$ intrinsic scatter in the residuals.}
\end{figure*}



\section{Summary}

We have argued that there is a sparsity of field and group galaxies in stellar population studies. Many predictions of the hierarchical clustering models compare the properties of field, cluster and group galaxies, however most of the field and group galaxy data available to date tend to be of lower quality than the cluster data. We hope that this work and its high quality data will serve as a step to improve the present state of knowledge based on early-type galaxies from low-density environments. 
 
We carried out an observationally homogeneous survey of 86 local early-type galaxies of mainly group (65), field (10) and isolated (8) objects with the aim of determining nuclear parameters, kinematic (this paper), age and metallicity gradients (subsequent paper). 

It is important to emphasize that our survey has the advantage of covering a large wavelength range,
from $\sim$ 3850 \AA\ to 6700 \AA, giving H$\alpha$ 
information which is important for emission correction (Section 4.2).

Most of the galaxies in the sample do not show 
obvious signs of disturbances nor tidal features in their
morphologies, although 11 galaxies belong to the Arp catalogue of 
peculiar galaxies (Arp 1966), of which only three (NGC~750, NGC~751 and 
NGC~3226) seem to be strongly interacting. Of the 86 galaxies with S/N $\geq$ 15 (per resolution element, for r$_e$/8 aperture), 57 had H$\beta$-index corrected for emission, the average correction was +0.190\AA\ in H$\beta$; 42 galaxies have \Oiii$\lambda$5007 emission correction, of which only 16 also show H$\alpha$ emission.

Our data allowed us to apply a better method for emission correction of 
the Balmer-line indices than the use of the uncertain 
 0.6$\times$\Oiii estimation. The new correction uses the 
intensity and equivalent width of the H$\alpha$ index (defined in Table 6).
We note that nebular emission could still be affecting the ages and 
metallicities derived from most of the data in the literature.

A central aspect of this work is that the determination of the errors in all of the measured and derived parameters is based on the analysis of the distribution of repeated observations. In this paper we are presenting data which have the advantage of several repeated observations (measurements) per galaxy. On average, to each galaxy we have eight corresponding spectral frames. 

Our main findings are that the index-$\sigma_0$ relations presented for low-density regions are not significantly different from those of cluster E/S0s. The slope of the index-$\sigma_0$ relations shown in Section 7 does not seem to change for early-type galaxies of different environmental densities, but the scatter of the relations seems larger for group, field and isolated galaxies than for cluster galaxies.

A thorough analysis of the highest signal-to-noise galaxies with discussions on the age and metallicity determinations is presented in a forthcoming paper.

\section*{Acknowledgements}

GD would like to thank CNPq-Brazil for the PhD fellowship, INAOE 
for hospitality during the visits to Mexico, Selwyn College and Cambridge
Philosophical Society for financial support.
The authors are grateful to the INAOE Committee for telescope time for 
supporting this project for five consecutive semesters and to 
the staff at Cananea for cheerful logistic help during the 
observations. We thank Am\^ancio Fria\c ca and Alessandro Bressan for 
several interesting discussions, Max Pettini and Alfonso Arag\'on-Salamanca 
for substantial comments to improve this work, and Ewan O'Sullivan for 
providing us 
with the details of the isolated sample selection previous to publication. 
We also thank an anonymous referee for a very thorough analysis of the 
manuscript. 
RJT and ET acknowledge the Mexican Research Council Project grants 40018-A-1 
and E32186, respectively.   



\appendix


\section{Sample complementary information}

  Table A1 contains complementary information about the galaxy sample, including environment information and our classification as group, field and isolated galaxy.

\begin{table*}
\label{TableA1}
\centering
\begin{minipage}{160mm}
\caption{Complementary details of the galaxy sample.}
\begin{tabular}{@{}lrrrrrrl@{}}
\hline
\multicolumn{1}{c}{Name} &  
\multicolumn{1}{c}{Type}  &
\multicolumn{1}{c}{V$_{rot}$} &
\multicolumn{1}{c}{r$_e$} & 
\multicolumn{1}{c}{log R$_{25}$} &
\multicolumn{1}{c}{PA$_{maj}$} &
\multicolumn{1}{c}{FP$_{res}$} &
\multicolumn{1}{c}{Environment information} \\
\multicolumn{1}{c}{} & 
\multicolumn{1}{c}{} & 
\multicolumn{1}{c}{(km/s)} &
\multicolumn{1}{c}{('')}&
\multicolumn{1}{c}{} &
\multicolumn{1}{c}{($^o$)} &
\multicolumn{1}{c}{} &
\multicolumn{1}{c}{} \\
\hline

ESO462-G015 & -5 & - & 21& 0.11& 166 &-0.05&{\bf Isolated}\\ 
MCG-01-03-018 & -3 &- &23* &0.00 & &0.29&{\bf Isolated}\\
NGC~0016 & -3 & 160& 28* & 0.27 &16& 0.02&{\bf Field}\\
NGC~0221 & -6 & 46 & 36 &0.13 & 170&0.19&M32, dwarf companion of M31, {\bf group} (LGG 11)\\
NGC~0315 & -4 & 32 & 37 & 0.20 &40 &0.09&{\bf Group} (LGG 14)\\
NGC~0474 & -2 & - & 34& 0.05 & 75&0.11 &Arp227, {\bf group} (LGG 20)\\ 
NGC~0584 & -5 & 157 &25 &0.26 & 55&-0.28&{\bf Group} (LGG 27)\\
NGC~0720 & -5 & 48 & 36& 0.29 &140 &0.02& {\bf Group} (LGG 38)\\
NGC~0750 & -5 & 40 &27* &0.10 &  &0.04&Arp166, pair with NGC~0751, {\bf group} (LGG 42)\\ 
NGC~0751 & -5 & - & 22*& 0.00& &0.20&Arp166, pair with NGC~0750, {\bf group} (LGG 42)\\
NGC~0777 & -5 & 53 & 34&0.08 & 155&0.15&{\bf Group} (LGG 42)\\
NGC~0821 & -5 & 89 & 50& 0.20 & 25&0.28&{\bf Isolated}\\
NGC~0890 & -3 & - & 34 & 0.16 &50* &-0.17&{\bf Field}\\
NGC~1045 & -3 & -&36* &0.28 & 40&0.19&{\bf Isolated}\\
NGC~1052 & -5 & 101 &34 &0.16&120&0.32&{\bf Group} (LGG 71)\\
NGC~1132 & -4.5 & - & 34&0.27& 140&0.06&{\bf Isolated}\\
NGC~1407 & -5 & 30 &70 &0.03&35&0.21&Eridanus {\bf group} (LGG 100)\\
NGC~1453 & -5 & - & 25&0.09 & &0.29& {\bf Group} (LGG 103)\\
NGC~1600 & -5 & 4 &45 &0.17 &15 &0.20&{\bf Field}\\
NGC~1700 & -5 & 75 &18 & 0.20 &90 &-0.51&{\bf Group} (LGG 123)\\
NGC~1726 & -2 & 40 & 24& 0.12 & 170&-0.10&{\bf Field}\\
NGC~2128 & -3 & - & 23*&0.13 &60 &0.51&{\bf Isolated}\\ 
NGC~2300 & -2 & 6  & 31 & 0.14 &  &0.44&Arp114, {\bf group} (LGG 145)\\
NGC~2418 & -5 & 142  &  28* & 0.00 &  &0.31&Arp165, {\bf field}\\
NGC~2513 & -5 & 53 &33 & 0.09&170 &0.19&{\bf Field}\\
NGC~2549 & -2 & 115 & 17 & 0.48 &177 &-0.03&{\bf Field}\\
NGC~2768 & -5 & 78 & 64&0.28 &95&-0.06 &{\bf Group} (LGG 167)\\
NGC~2872 & -5 & 75 & 18& 0.06&22 &0.17&Arp307, {\bf field}\\
NGC~2911 & -2 & -  & 51&0.11&140&0.17&Arp232, {\bf group} (LGG 177) \\
NGC~2974 & -5 & 202 & 24&0.23& 40&0.21&{\bf Group} (LGG 179) \\
NGC~3091 & -5 & 71 & 33& 0.20&149&0.27&{\bf Group} (LGG 186)\\
NGC~3098 & -2 & 130  & 15&0.57&90 &-0.12&{\bf Field}\\
NGC~3115 & -3 & 273 & 32 & 0.47 &40&-0.11&{\bf Field}\\
NGC~3139 & -2 & - &22* & 0.08&  &1.46&{\bf Field}\\
NGC~3156 & -2 & 79  & 14& 0.24&47 &-0.54 &{\bf Group} (LGG 192)\\
NGC~3193 & -5 & 80 &27 &0.05 & &0.26&Arp316, {\bf group} (LGG 194)\\
NGC~3226 & -5 & 40 & 34 &0.05 &15 &0.40&Arp094, {\bf group} (LGG 194)\\
NGC~3245 & -2 & - & 27 & 0.26 & 177&0.09&{\bf Group} (LGG 197)\\
NGC~3377 & -5 & 86 &34 &0.24 &35 & 0.03&Leo {\bf group} (LGG 217)\\
NGC~3379 & -5 & 53  &  35 & 0.05  & & -0.18 &Leo {\bf group} (LGG 217)\\
NGC~3384 & -3 & -  & 25 & 0.34 &53 &-0.23 &Leo {\bf group} (LGG 217)\\
NGC~3412 & -2 & - & 26 & 0.25 &155 &-0.38 &Leo {\bf group} (LGG 217)\\
NGC~3414 & -2 & 63 & 21 & 0.14 & &0.18 &Arp162, {\bf group} (LGG 227)\\
NGC~3599 & -2 & -  & 24&0.10 & & -0.12&{\bf Field}\\
NGC~3607 & -2 & 108  & 43  & 0.30 &120 &0.22&{\bf Field} \\
NGC~3608 & -5 & 26  & 34 & 0.09&75 &0.27&{\bf Group} (LGG 237)\\
NGC~3610 & -5 & 143 & 15 & 0.07 & &-0.50&{\bf Group} (LGG 234)\\
NGC~3613 & -5 & 141 & 27& 0.32& 102&-0.04 &{\bf Group} (LGG 232)\\
NGC~3636 & -5 & - &20* &0.00& &-0.04&{\bf Group} (LGG 235)\\
NGC~3640 & -5 & 114  & 32& 0.10&100&0.00&{\bf Group} (LGG 233)\\
NGC~3665 & -2 & 103 & 29 & 0.08 &30 &0.00 &{\bf Group} (LGG 236)\\
NGC~3923 & -5 & - & 50&0.18 & 50&0.01 &{\bf Group} (LGG 255)\\
NGC~3941 & -2 & - & 23 & 0.18 &10 &-0.39&{\bf Field}\\
NGC~4125 & -5 & 97 & 58&0.26&95&0.05&{\bf Group} (LGG 274)\\
NGC~4261 & -5 & 69 & 36 & 0.05&160 &0.13 &3C270, {\bf group} (LGG 278)\\
NGC~4365 & -5 & 43 & 50&0.14 &40 & 0.04&Virgo, {\bf group} (LGG 289)\\
NGC~4374 & -5 & 23 & 51& 0.06 &135 & 0.08&Virgo, {\bf group} (LGG 292)\\
NGC~4494 & -5 & 67  &  49& 0.13 & &-0.33 &{\bf Group} (LGG 294) \\
NGC~4550 & -1.5 & 122  & 15 & 0.55 &178 &-0.03&{\bf Field}\\
NGC~4754 & -3 & 89 & 26 & 0.27 & 23&-0.08 &Virgo, {\bf group} (LGG 289)\\
NGC~5322 & -5 & 20  & 34& 0.18 & 95&-0.07& {\bf Group} (LGG 360)\\
NGC~5353 & -2 & - & 15 & 0.30 &145 &0.03&{\bf Group} (LGG 363) \\
NGC~5354 & -2 & - & 18&0.04 & & -0.14&{\bf Group} (LGG 361) \\

\hline
\end{tabular}
\end{minipage}
\end{table*}

\setcounter{table}{0}
\begin{table*}
\label{TableA1}
\centering
\begin{minipage}{160mm}
\caption{Continued.}
\begin{tabular}{@{}lrrrrrrl@{}}
\hline
\multicolumn{1}{c}{Name} &  
\multicolumn{1}{c}{Type}  &
\multicolumn{1}{c}{V$_{rot}$} &
\multicolumn{1}{c}{r$_e$} & 
\multicolumn{1}{c}{log R$_{25}$} &
\multicolumn{1}{c}{PA$_{maj}$} &
\multicolumn{1}{c}{FP$_{res}$} &
\multicolumn{1}{c}{Environment information} \\
\multicolumn{1}{c}{} & 
\multicolumn{1}{c}{} & 
\multicolumn{1}{c}{(km/s)} &
\multicolumn{1}{c}{('')}&
\multicolumn{1}{c}{} &
\multicolumn{1}{c}{($^o$)} &
\multicolumn{1}{c}{} &
\multicolumn{1}{c}{} \\
\hline

NGC~5363 & 90.0  & - & 36& 0.19 &135 &0.06 &{\bf Group} (LGG 362)\\
NGC~5444 & -4 & - & 27&0.06 &90&0.17 & {\bf Group} (LGG 370)\\
NGC~5557 & -5 & - & 30& 0.09&105 &-0.04& {\bf Group} (LGG 378)\\
NGC~5576 & -5 & 14 & 18&0.20 & 95&-0.14&{\bf Group} (LGG 379)\\
NGC~5638 & -5 & 62 & 28 & 0.05& 150& -0.01&{\bf Group} (LGG 386)\\
NGC~5812 & -5 & 40 & 26& 0.06& &0.24&{\bf Field}\\ 
NGC~5813 & -5 & 36 & 57& 0.14 &145 &0.23&{\bf Group} (LGG 393)\\
NGC~5831 & -5 & 27 &26 & 0.04&55 & 0.24&{\bf Group} (LGG 393)\\
NGC~5845 & -4.6 & 127 &12* &0.19 &150 &0.81&{\bf Group} (LGG 392)\\
NGC~5846 & -5 & 4  & 62  & 0.03  & &0.12&Pair with NGC~5846A, {\bf group} (LGG 393)\\
NGC~5846 A& -6  & 74 & 6* & 0.15  &120 & 0.35 &Pair with NGC~5846, {\bf group} (LGG 393)\\
NGC~5854 & -1 & - & 15 & 0.54 &55 &-0.24&{\bf Group} (LGG 393)\\
NGC~5864 & -2  & - & 44* & 0.49 &68 &0.24&{\bf Group} (LGG 393)\\
NGC~5869 & -2 & - & 36*&0.14 &125 &0.32 &{\bf Group} (LGG 393)\\
NGC~5982 & -5 & 45 & 24& 0.12&110 &0.00&{\bf Group} (LGG 402)\\
NGC~6172 & -4 & -  & 16* & 0.00 & &-0.12&{\bf Isolated}\\
NGC~6411 & -5 & 14 & 29& 0.10&70&-0.18 &{\bf Isolated}\\
NGC~7302 & -3 & - & 13& 0.21& 100* &0.15&{\bf Field}\\
NGC~7332 & -2 & 134 & 15& 0.56 &155 &-0.30&{\bf Field}\\
NGC~7454 & -5 & 23 & 25 & 0.15&150 &-0.07&{\bf Group} (LGG 469)\\
NGC~7585 & -1 & -& 24& 0.07&105 &-0.09&Arp223, {\bf field}\\
NGC~7619 & -5 & 69 & 37& 0.04& 30&0.29&Pegasus I {\bf group} (LGG 473) \\
NGC~7626 & -5 & 20 & 39& 0.05& &0.25&Pegasus I {\bf group} (LGG 473) \\

\hline
\end{tabular}

References: Morphologies are from the {\it Third Reference Catalogue} (RC3). The {\it maximum} rotation velocity V$_{rot}$ is from Prugniel \& Simien (1996). Effective radius r$_e$ and the R$_{25}$ parameter are from Trager \etal (2000b) and RC3 catalog. The Fundamental Plane residual FP$_{res}$ was derived from eq.(4) in Prugniel \& Simien (1996). The asterisk symbols assign estimates made by the authors.  Environmental information is from E. O'Sullivan (private communication); the group identification was extracted from Garcia 1993 (LGG group number in parenthesis). 
\end{minipage}
\end{table*}



\section{Fully corrected line-strength indices}

We present the final corrected central r$_e$/8 index measurements and associated 
errors for our galaxy sample. The Lick indices presented in Table B.1 were calibrated to the Lick/IDS system, and corrected for velocity dispersion and nebular emission (in this case mainly the Balmer lines).

\begin{table*}
\centering
\vspace{23.5 cm}
\caption{Fully corrected Lick/IDS indices for the central r$_e$/8 extraction.}
\includegraphics{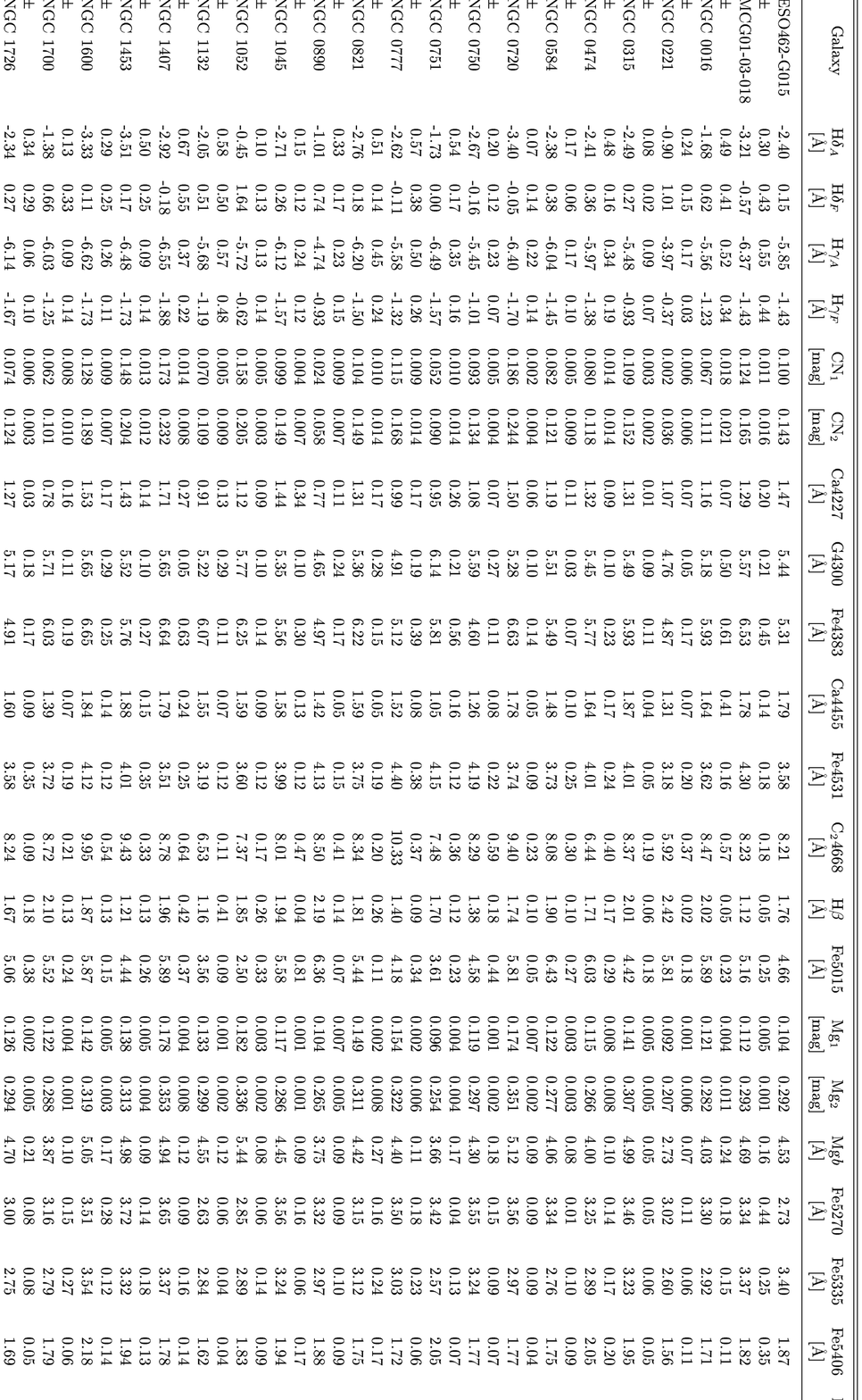}
\end{table*}
\setcounter{table}{0}
\begin{table*}
\centering
\vspace{23.5 cm}
\caption{Fully corrected Lick/IDS indices for the central r$_e$/8 extraction: continued. }
\includegraphics{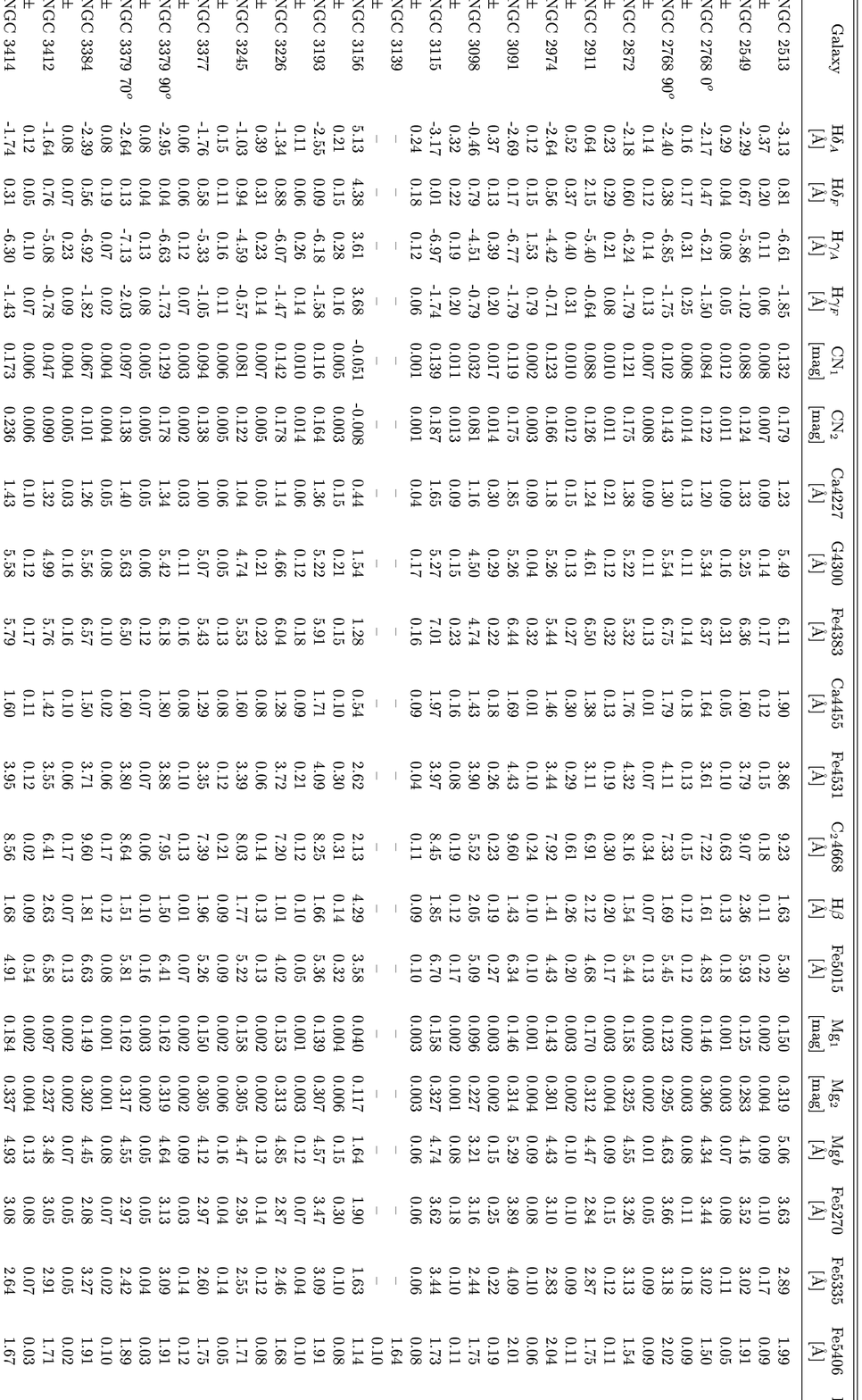} 
\end{table*}
\setcounter{table}{0}
\begin{table*}
\centering
\vspace{23.5 cm}
\caption{Fully corrected Lick/IDS indices for the central r$_e$/8 extraction: continued. }
\includegraphics{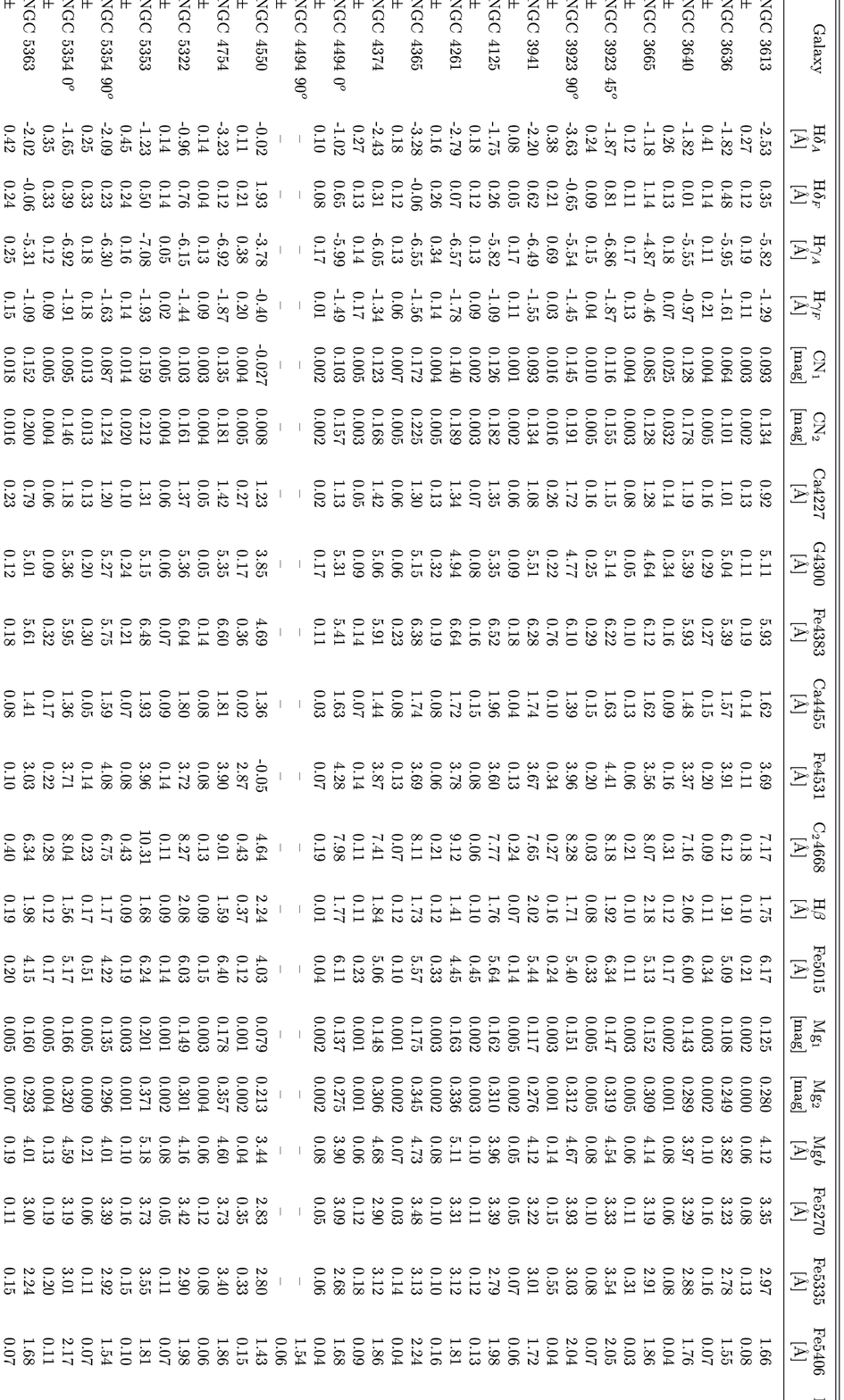}
\end{table*}
\setcounter{table}{0}
\begin{table*}
\centering
\vspace{23.5 cm}
\caption{Fully corrected Lick/IDS indices for the central r$_e$/8 extraction: continued.  }
\includegraphics{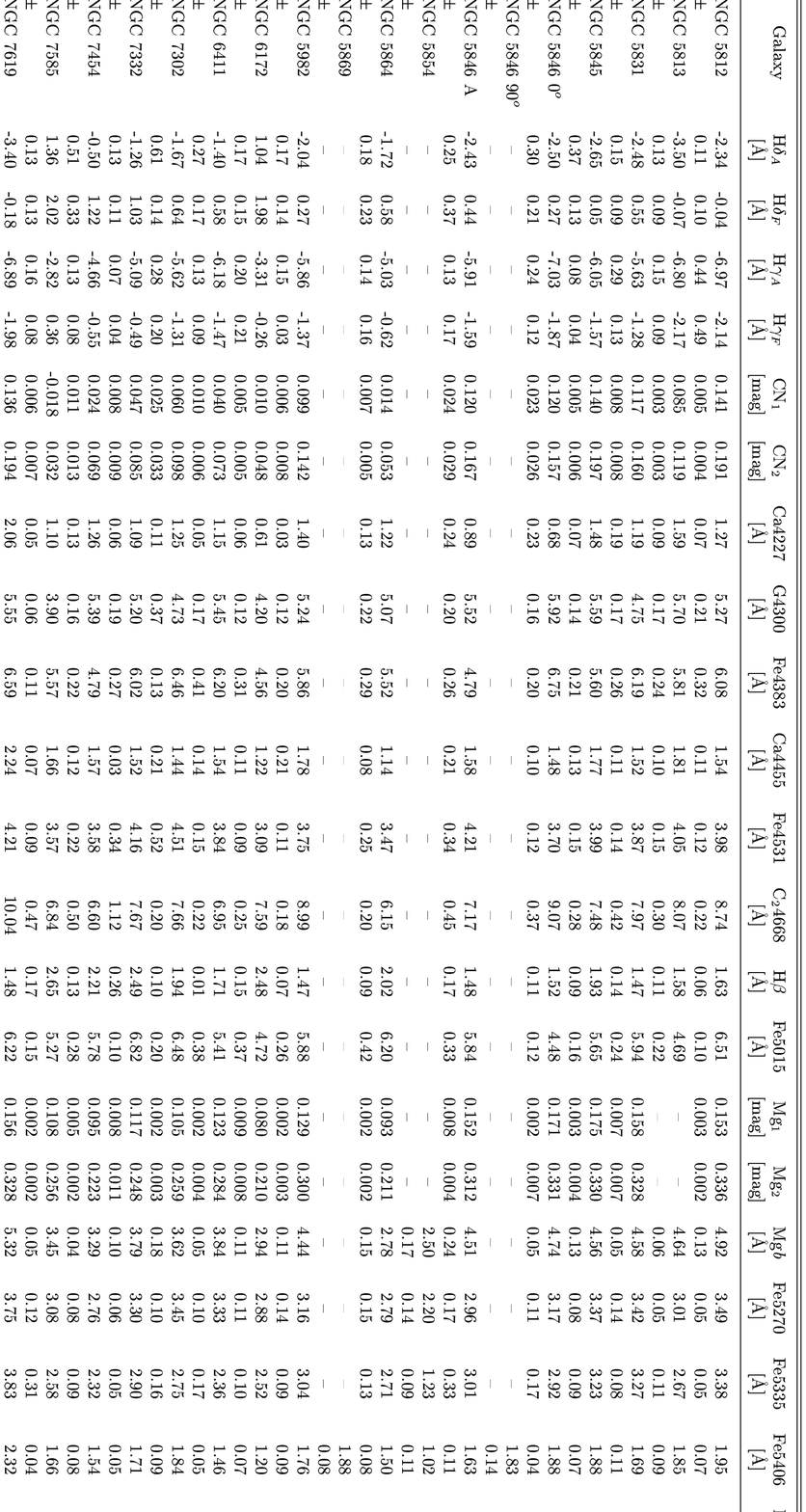}
\end{table*}


\label{lastpage}
\end{document}